\documentclass[DIV12]{scrartcl}
\usepackage[utf8]{inputenc}
\usepackage{microtype}
\DeclareMathAlphabet{\mathcal}{OMS}{cmsy}{m}{n}

\usepackage{amsthm,amssymb,amsmath}
\usepackage{xspace}
\usepackage{paralist}
\usepackage[hidelinks]{hyperref}

\newcommand{\true}{\top}
\newcommand{\false}{\bot}

\newcommand{\card}[1]{\left|#1\right|}

\newcommand{\kw}[1]{{\mathsf{#1}}\xspace}

\newcommand{\wellformed}{\kw{wf}}

\newcommand{\wf}{\wellformed}

\newcommand{\ineig}[2]{\kw{IN}^{#1}(#2)}
\newcommand{\ineigtwo}[2]{\kw{IN}_2^{#1}(#2)}

\newcommand{\myparagraph}[1]{\paragraph{#1.}}

\newcommand{\calA}{\mathcal{A}}
\newcommand{\calB}{\mathcal{B}}

\newcommand{\calD}{\mathcal{D}}

\newcommand{\calG}{\mathcal{G}}

\newcommand{\calI}{\mathcal{I}}
\newcommand{\calJ}{\mathcal{J}}

\newcommand{\calW}{\mathcal{W}}
\newcommand{\calX}{\mathcal{X}}

\newcommand{\shred}{\mathrm{SHR}}
\newcommand{\treeif}{\mathrm{TR}}

\newcommand{\gaifman}[1]{\calG(#1)}

\newcommand{\NN}{\mathbb{N}}

\newcommand{\dclass}[1]{\kw{#1}}
\newcommand{\gctwo}{\dclass{GC}^{\smash{2}}}

\newcommand{\gftwo}{\dclass{GF}^2}

\newcommand{\uidclass}{\dclass{ID}[1]}
\newcommand{\bidclass}{\dclass{ID}[2]}
\newcommand{\idclass}{\dclass{ID}}
\newcommand{\foneldclass}{\dclass{FR[1]}}
\newcommand{\fone}{\foneldclass}
\makeatletter
\newcommand{\raisemath}[1]{\mathpalette{\raisem@th{#1}}}
\newcommand{\raisem@th}[3]{\raisebox{#1}{$#2#3$}}
\makeatother
\newcommand{\myfone}[1]{\fone^{\raisemath{-2.43pt}{\smash{\mathrm{#1}}}}}
\newcommand{\myfonesafe}[1]{\fone^{\mathrm{#1}}}
\newcommand{\foneha}{\myfone{Hnl}}
\newcommand{\fonehba}{\myfone{Fnl}}
\newcommand{\fonehbasafe}{\myfonesafe{Fnl}}
\newcommand{\fonesh}{\myfone{SH}}

\newcommand{\stt}{\dclass{S2T}}

\newcommand{\ufdclass}{\dclass{UFD}}

\DeclareMathOperator{\funct}{funct}

\DeclareMathOperator{\dom}{dom}

\newcommand{\entunr}{\models}

\newcommand\restr[2]{{
  \kern-\nulldelimiterspace 
  #1 
  _{|#2} 
  }}

\newcommand{\quot}[2]{#1/{#2}}

\newcommand{\alcqib}{\mathcal{ALCQI}b}
\newcommand{\alcf}{\mathcal{ALCF}}
\newcommand{\dlr}{\mathcal{DLR}_{reg}}

\newcommand{\arity}[1]{\card{#1}}

\newcommand{\class}{\mathsf{CL}}

\newcommand{\elt}{\mathrm{Elt}}

\renewcommand{\a}{\mathrm{a}}

\renewcommand{\r}{\mathrm{r}}

\newcommand{\sigmas}{\sigma_{\mathrm{S}}} 

\renewcommand{\S}{\mathrm{S}}
\newcommand{\T}{\mathrm{T}}

\newcommand{\sroiq}{\mathcal{SROIQ}}

\newcommand{\f}{\mathrm{f}}

\newcommand{\deft}[1]{\emph{#1}}

\newtheorem{theorem}{Theorem}[section]
\newtheorem{example}[theorem]{Example}
\newtheorem{lemma}[theorem]{Lemma}
\newtheorem{proposition}[theorem]{Proposition}

\newtheorem{corollary}[theorem]{Corollary}
\newtheorem{definition}[theorem]{Definition}

\makeatletter
\newcommand*{\defeq}{\mathrel{\rlap{%
  \raisebox{0.3ex}{$\m@th\cdot$}}%
  \raisebox{-0.3ex}{$\m@th\cdot$}}%
  =}
\makeatother

\title{Combining  Existential Rules\\and Description Logics\\(Extended Version)}

\author{
\begin{tabular}[t]{c} 
Antoine Amarilli \\
{\normalfont T\'el\'ecom ParisTech; Institut Mines--T\'el\'ecom; CNRS LTCI} \\
{\normalfont antoine.amarilli@telecom-paristech.fr} \\[0.5em]
Michael Benedikt \\
{\normalfont University of Oxford} \\ 
{\normalfont michael.benedikt@cs.ox.ac.uk}
\end{tabular}
}

\begin{document}

\maketitle

\begin{abstract}
  Query answering under \emph{existential
  rules} --- implications
with existential quantifiers in the head --- is known to be decidable when imposing restrictions on the
rule bodies 
such as 
 frontier-guardedness
\cite{baget2010walking,baget2011rules}.
Query answering is also decidable for \emph{description logics}
\cite{baader2003description}, which further
allow disjunction and \emph{functionality constraints} (assert that certain
relations are functions); however, they
are focused on ER-type schemas,
where relations have arity two.

This work investigates how to  get the best of both worlds: having decidable existential rules
on arbitrary arity relations, while allowing rich description logics, including functionality constraints, on
arity-two relations. We first show negative results on combining such decidable
languages. Second, we introduce an expressive set of existential rules
(frontier-one rules with a certain restriction) which 
can be combined with 
powerful  constraints on arity-two relations (e.g. $\gctwo, \alcqib$) while
retaining decidable query answering. Further,
we provide conditions to add functionality constraints on the
higher-arity relations.
\end{abstract}

\section{Introduction}
\label{sec:intro}
Recent years have seen an explosion of techniques for solving the \emph{query answering problem}:
given a query $q$, a conjunction~$F$ of atoms, and a set of logical constraints $\Sigma$,
determine whether $q$ follows
from  $F$ and $\Sigma$. In databases this is called \emph{querying under
constraints} or the \emph{certain answer problem}, seeing
$F$ as an incomplete database, and $\Sigma$ as restrictions on the possible
completions. 
For researchers working on description logics, $F$ is referred to as
the \emph{A-box} and  $\Sigma$ the \emph{T-box}.
In both communities
$q$ is usually a \emph{conjunctive query},  
an existential quantification of conjunctions of atoms,
equivalent to a basic SQL SELECT.
We will make this assumption  throughout this work, referring for simplicity to the problem 
as just ``query answering'' (QA).

QA 
is undecidable when $\Sigma$ ranges over arbitrary
first-order logic constraints. This motivates
the search for restricted constraint languages with decidable
QA.
Within the description logic community, powerful such languages
were developed to express constraints on vocabularies of arity two. The unary relations are referred
to as \emph{concepts} while the binary ones are the \emph{roles}. The languages
can build
new concepts and roles from basic ones via Boolean operations and (limited)
quantification, and many of them, such as DL-Lite~\cite{calvanese2005dl} or
$\alcqib$~\cite{tobies2001complexity},
may
restrict the input roles $R(x,y)$ to be \emph{functional} -- for all $x$ there
is at most one $y$ such that $R(x, y)$.
Functionality constraints are crucial to faithfully model many real-world
relationships: the relationship of a person
to their birthdate, the relationship of an event to its starting time, etc.
Hence, description logics are very powerful languages for \emph{arity-two vocabularies}.

In  parallel, the AI and database  communities have developed rich constraint languages
on arbitrary
arity via \emph{existential rules} or \emph{tuple-generating dependencies} (TGDs). Existential rules are constraints of the form
$\forall \mathbf x ~ (\phi(\mathbf x) \rightarrow \exists \mathbf y ~
\psi(\mathbf{x'}, \mathbf y))$
where $\mathbf{x'} \subseteq \mathbf{x}$ and $\phi$ and $\psi$ are conjunctions of atoms.
They generalize the well-known \emph{inclusion dependencies} or \emph{referential constraints} in databases \cite{AHV}, and
can also express mapping relationships used in data exchange \cite{dataex} and data integration \cite{dataint}.
Although QA over general rules is undecidable, important subclasses are decidable. 
First, decidability holds whenever the \emph{chase} procedure \cite{AHV} is
guaranteed to terminate, which is ensured by a number of conditions on the rules,
 e.g., weak 
acyclicity \cite{dataex},  joint acyclicity \cite{jointlyacyclic},  or the very restricted class of
source-to-target TGDs.
See \cite{acyclicitydl} for a survey and \cite{baget2014extending} for a recent
study.
A second class of tame constraints are those that admit bounded-treewidth models.
There are several such classes, such
as \emph{guarded TGDs} \cite{cali2012general}, \emph{frontier-guarded TGDs}
\cite{baget2010walking}, or the more general \emph{greedy bounded-treewidth
sets} \cite{baget2011walking}.
However,
many
features of description logics, such as
disjunction or functionality restrictions, cannot be expressed by existential
rules.

Could we then enjoy the best of both worlds, by allowing both description logic
constraints and existential rules, while maintaining the decidability of QA?
This paper studies to what extent both paradigms can be combined, by
looking for classes of constraints
with  decidable QA over relational schemas of \emph{arbitrary arity} that can 
\begin{inparaenum} 
\item express non-trivial existential rules over any relation in the schema and 
\item assert expressive constraints (e.g., in $\alcqib$) on the \emph{arity-two subschema} --- the subset
of the relations of arity one and two within the schema
\end{inparaenum}

Our first results (Section~\ref{sec:undecide}) are negative: we show that arity-two languages featuring
functionality constraints on the arity-two subschema may lead to undecidable QA when
combined with even very simple acyclic rules (source-to-target TGDs, $\stt$), or
with the simplest existential
rules that export two variables (frontier-two inclusion dependencies,
$\bidclass$).
More surprisingly, undecidability can occur 
with rules exporting
only a single variable,  the class of 
\emph{frontier-one dependencies} $\foneldclass$
of
\cite{baget2009extending}.
We say the  existential rule languages $\stt$, $\bidclass$, $\foneldclass$   are \emph{destructive of arity-two QA}.

We then show (Section~\ref{sec:decide})
that by restricting $\foneldclass$ slightly, imposing that the head of the rules
have a certain tree shape (denoted ``non-looping''),  we can obtain a
  class of existential rules 
that 
can be combined
with expressive constraints on the arity-two schema while
maintaining decidable QA (we call this \emph{not destructive}). 
The reduction proceeds in two steps.
We first handle rules with tree-shaped bodies, via a direct rewriting technique to
constraints on an arity-two encoding of the schema.
Second, we handle rules with non-tree-shaped bodies,
showing that the bodies can be soundly replaced  by a tree-shaped
approximation. 
Soundness is proven by extending the technique of ``treeification'' used
previously in many modal and guarded logics (e.g., \cite{bgo}), showing that models of the constraints
can be ``unraveled'' to be tree-shaped.

We go on to study (Section~\ref{sec:fds}) the addition of
functional dependencies (FDs), a well-known generalization of description logic functionality
constraints to arbitrary arity. QA with existential rules and FDs is generally
undecidable 
unless their interaction with the existential rules is controlled, e.g., by
imposing the non-conflicting condition \cite{cali2012towards}. We show that FDs
can be added to our existential rules while maintaining decidable QA with the
arity-two constraints, as long as the
non-conflicting condition is satisfied. 
As in the standard
non-conflicting setting, we show that the FDs can
always be satisfied unless the initial facts violate them.
We prove this by modifying the unraveling argument.

Our results have the advantage that QA for our combined constraints reduces
to QA on an arity-two schema; hence, 
existing QA algorithms for rich description logics could be extended to
arbitrary arity signatures with expressive constraints. 

\myparagraph{Related work}
A great deal of research has centered around the integration of DLs with
Datalog-style rules, including work as early as the 1990's, when  the 
languages AL-Log \cite{allogconf}
and CARIN \cite{carinj} were introduced.
 AL-Log  links Horn 
rules with concepts from a description logic terminology,
while  the later language CARIN provides a broader framework
allowing both concepts and roles from a terminology to appear
in rules.  \cite{carinj} provides both entailment algorithms for CARIN
and undecidability results exploring the borderline for combining
rules and DLs.

Datalog rules, however, unlike the existential rules that we consider in this
work, do not allow existential quantification in the head, so they cannot assert the
existence of higher-arity facts on fresh elements.

Another approach to combination are
  description logics
 that
support  higher-arity relations directly. Languages such as $\dlr$ \cite{calvanese2008conjunctive}
give some support for higher arity while retaining a DL-style syntax.
Unlike them, we support existential rules with cyclic bodies that cannot be
encoded in $\dlr$, as well as arbitrary higher-arity functional
dependencies that go beyond DL-expressible functionality assertions. On the other hand, we do not support some features of $\dlr$,
such as regular expression on role paths. Indeed, we do not consider
the interaction of rules with DLs supporting transitivity and other recursion mechanisms 
\cite{shiq},  focusing instead
only on
first-order-expressible constraints
given by decidable DLs  
 and existential
rules.

\section{Preliminaries}
\label{sec:prelim}
\myparagraph{Signatures, facts, queries} A \deft{signature}~$\sigma$ consists of \deft{relation names} (e.g. $R$) and an associated
\deft{arity} (e.g. $\arity{R}$). We write $\sigma$ as $\sigma_{\leq 2} \sqcup
\sigma_{>2}$, containing respectively the relations of arity $\leq 2$ and the
\deft{higher-arity} relations 
with arity $> 2$.
An \deft{atom}~$R(\mathbf{x})$ consists of a relation name $R$ and an
$\arity{R}$-tuple $\mathbf{x}$ of variables. 
A $\sigma$-\deft{fact} (or just \deft{fact} when
$\sigma$ is clear from context) is a conjunction
 of atoms using relations in $\sigma$.
A Boolean \deft{conjunctive
query} (or CQ) is an existentially quantified conjunction of atoms.
In this paper we assume for simplicity that CQs are Boolean, i.e., have no free
variables, and we disallow constants. This is without loss of generality:
for non-Boolean queries we can enumerate all possible assignments, and constants
can be encoded with fresh unary relations.

\myparagraph{Constraints, QA}
We consider constraints that are formulae in
function-free and constant-free
first-order logic (FO), on the signature $\sigma$. 
A $\sigma$-\deft{interpretation} $\calI$ (or just \deft{interpretation}) consists of
a \deft{domain} $\dom(\calI)$ and an \deft{interpretation function} $\cdot^{\calI}$ mapping each relation $R$ of $\sigma$ to a set $R^{\calI}$ of
$\arity{R}$-tuples of~$\dom(\calI)$. The definition of $\calI$ \deft{satisfying}
a FO formula~$\phi$, written $\calI \models \phi$, is standard. 
A \deft{witness} $\calW$ of $F$ in $\calI$ is an
interpretation that maps each relation $R$
to the tuples in $R^{\calI}$ obtained by substituting
the atoms of $F$ using some variable binding~$\mathbf{w}$ such that $\calI \models
F(\mathbf{w})$.

We study the
\deft{query answering} problem (QA): given a fact~$F$, a set of constraints $\Sigma$, and a CQ $q$,
decide the validity of $\forall \mathbf{x} ~ (F(\mathbf{x}) \wedge \Sigma \rightarrow
q)$;
that is, whether  $F$
and $\Sigma$ entail $q$. In this case, we write $F \wedge \Sigma \entunr q$.
The \emph{combined complexity} of
QA, for a fixed class of
constraints, is the complexity of
deciding it when all of $F$, $\Sigma$ (in the constraint class) and $q$ are given as input. If we assume
that $\Sigma$ and $q$ are fixed, and only $F$ is given as input, then we define
instead the \deft{data complexity}.

The  QA problem above allows arbitrary FO constraint classes. Below we present
two kinds of integrity constraints that are known to enjoy decidable QA.

\myparagraph{Existential rules}
An \deft{existential
rule} (or \deft{tuple-generating dependency}, or TGD) is a logical constraint of the
form $\forall \mathbf{x} ~ (\phi(\mathbf{x}) \rightarrow \exists \mathbf{y} ~
\psi(\mathbf{x'}, \mathbf{y}))$, with $\mathbf{x'} \subseteq \mathbf{x}$, where the \deft{body} $\phi$ and \deft{head}
$\psi$ are conjunctions of atoms. Equality atoms and constants are disallowed.
For brevity, in rules we often omit the quantification on
$\mathbf{x}$ and write `$\wedge$' as a comma.  A rule is \deft{single-head} if its head
consists of only one atom.

QA is undecidable for general rules (following from \cite{beeri1981implication}).
One class of rules with decidable QA are those satisfying \emph{acyclicity}
conditions. 
We will show negative results for one of the most restrictive classes,
 the class $\stt$ of \emph{source-to-target TGDs}, where
$\sigma$ is partitioned as $\sigma = \sigma_{\S} \sqcup \sigma_{\T}$, the bodies of all rules
only use relations in~$\sigma_{\S}$, and the heads only use relations in~$\sigma_{\T}$. Our results on $\stt$ extend to more permissive acyclicity
conditions, such as those mentioned in the introduction.

A second class of decidable rules guarantees that it suffices to consider
bounded-treewidth interpretations, usually because of constraints on the rule bodies. 
We focus on the class $\foneldclass$ of \deft{frontier-one rules},
following \cite{baget2009extending}: the 
\deft{frontier} of a rule is the set $\mathbf{x'}$ of variables that occur both
in the body and the head, and a rule is frontier-one if $\card{\mathbf{x'}} =
1$.
The class of \deft{inclusion dependencies} $\idclass$  
imposes that 
the head and body are single atoms
where each variable is used only once and that the frontier is not empty, and
we will focus on the class $\bidclass$ of
the inclusion dependencies with frontier size~$2$.
QA is decidable for $\foneldclass$ \cite{baget2009extending}.
For $\idclass$
it is decidable and has PTIME data complexity 
\cite{cali2003query}.

Existential rules can be augmented with \deft{functional dependencies}
(FDs), which are variants of
existential rules that impose equalities. Writing $\forall \mathbf{x} = \forall
x_1 \cdots \forall x_n$ and similarly for~$\mathbf{y}$, an FD on the relation $R$ is of the
form:\\[.1cm]
\mbox{$\forall \mathbf{x} \mathbf{y} ~ (R(x_1, \ldots, x_n) \wedge R(y_1,
\ldots, y_n) \wedge \bigwedge_{l \in L} x_l = y_l) \rightarrow x_r = y_r$}\\[.1cm]
for some $1 \leq r \leq \arity{R}$ and some subset $L \subseteq \{1, \ldots,
\arity{R}\}$
which we call the \deft{determiner} of the FD.
QA is undecidable when combining existential rules and arbitrary FDs, for
instance it is undecidable for $\bidclass$ and FDs \cite{cali2003decidability}.

\myparagraph{Arity-two constraints}
The second kind of tame constraints are \deft{arity-two constraints},
which are \emph{only defined on~$\sigma_{\leq 2}$}. The most general such
language that we study is the \deft{two-variable guarded
fragment with counting quantifiers}, $\gctwo$ \cite{kazakov2004polynomial}.
$\gctwo$  is the smallest class
of constant-free FO formulae with at most two variables, containing all atoms for
$\sigma_{\leq 2}$ relations,
closed under Boolean connectives,
under \emph{guarded} universal and existential quantification,
and under 
\emph{number quantifications}:
if $\phi(x, y)$ is a $\gctwo$
formula and $A(x, y)$ is an arity-two atom with two free variables (the
\emph{guard}),
then
$\exists^{\geq n} y ~ A(x, y) \wedge \phi(x, y)$
and $\exists^{< n} y ~ A(x, y) \wedge \phi(x, y)$
are formulae,  where $n$ is an 
integer.

QA for $\gctwo$ is decidable and its data complexity is in co-NP
\cite{pratt2009data}. 

\deft{Description logics} (DLs) are arity-two constraint languages. Examples of
DLs are DL-Lite~\cite{calvanese2005dl}, a lightweight DL often used in the context of
ontology-based data access, and $\alcqib$ \cite{tobies2001complexity}, a more expressive
DL  that can make full use of \emph{number restrictions}, a useful feature in practice.
Both DL-Lite and $\alcqib$ can assert \deft{concept inclusions} like $C \sqsubseteq C'$, where $C$
and $C'$ are \deft{concepts} (arity~$1$ relations), meaning that $C'$ holds
whenever $C$ does; and \deft{functionality assertions} $\funct(R)$, where $R$ is
a \deft{role} (an arity~$2$ relation), corresponding to
$\forall x ~ \exists^{\leq 1} y ~ R(x, y)$ in $\gctwo$, or
to the FD: $\forall x_1 x_2 y_1 y_2 ~ R(x_1, x_2) \wedge R(y_1,
y_2) \wedge x_1 = y_1 \rightarrow x_2 = y_2$.
Despite its expressiveness, $\alcqib$  can still, as DL-Lite, be           
captured by $\gctwo$, which
implies  decidable QA.

Roles and concepts can be \deft{atomic} (i.e., from $\sigma_{\leq 2}$) or defined using
constructors; we give some examples from $\alcqib$.
The \deft{inverse} $R^-$ of an atomic role $R$ is such that $R^-(b, a)$ holds whenever
$R(a, b)$ does. An \deft{intersection} of roles, which is written $R_1 \sqcap \cdots
\sqcap R_n$, holds for $(a, b)$ whenever $R_i(a, b)$ holds for all $1 \leq i
\leq n$. 
$\true$ and $\false$ are the \deft{true} and \deft{false} concepts.
The \deft{intersection} of concepts $C_1, \ldots, C_n$, written $C_1 \sqcap
\cdots \sqcap C_n$, holds whenever each of the $C_i$ does. The \deft{negation}
$\neg C$ of a concept $C$ holds for elements where $C$ does not hold. An
\deft{existential concept} $\exists R . C$ for a role $R$ and concept $C$ holds
for every element~$a$ such that $\exists b ~ R(a, b) \wedge C(b)$ does.
Note that many of these features (e.g., functionality assertions and negation)
cannot be expressed as existential rules.

\myparagraph{Combining constraint classes}
For any class $\class$ of existential rules, we call
$\class$
\deft{non-destructive} (of arity-two QA)
if QA is decidable for the class $\class \wedge
\gctwo$ of conjunctions of constraints of $\class$ (on $\sigma$) and
of constraints of $\gctwo$ (on $\sigma_{\leq 2}$).
Otherwise, we call $\class$ \deft{destructive}.

\section{Negative Results for Combination}
\label{sec:undecide}
We now present classes of existential rules which have decidable QA but are
destructive.
First, we observe that even the simplest class of rules
that ensures decidability based on chase termination, the class $\stt$ of
source-to-target TGDs,
is destructive. This is not so surprising,
since the arbitrary constraints on the arity-two signature
may add dependencies that are not source-to-target.

\begin{theorem}
  \label{thm:wadl}
  $\stt$ is destructive of arity-two QA, even when the whole $\sigma$ has arity
  two and there is no query 
  (i.e., this is just the \deft{satisfiability problem} asking whether the fact
  and constraints are satisfiable).
\end{theorem}

Thus we move on to classes of existential rules that are decidable because of
guardedness assumptions.

We first observe that the class $\bidclass$ of frontier-two inclusion dependencies is
destructive of arity-two QA. In fact, functionality assertions on the binary
relations are sufficient to get undecidability,
because they can be lifted to
functionality assertions on higher-arity relations using $\bidclass$. Thus, 
following a standard reduction
from QA to entailment of dependencies as in \cite{cali2003decidability},
we can use the undecidability of entailment for $\bidclass$ and FDs (Theorem~2
of \cite{mitchell1983implication}, which we adapt slightly) and prove the
following:

\begin{theorem}
  \label{thm:biddl}
  $\bidclass$ is destructive of arity-two QA. In particular,
  QA is undecidable for $\bidclass \wedge \calD$, for any DL $\calD$ (such as
  DL-Lite) featuring
  functionality assertions.
\end{theorem}

More surprisingly, frontier-one rules $\foneldclass$ are destructive of arity-two QA,
 even though they can only export a single variable, and this holds even when
 the whole $\sigma$ has arity two. The reason is that $\foneldclass$ may be more
 expressive than $\gctwo$ as it can disobey the two-variable restriction.

\begin{theorem}
  \label{thm:fronedl}
    $\foneldclass$ is destructive of arity-two QA, even when the whole $\sigma$
    has arity two and there is no query.
\end{theorem}

This motivates the search for more restricted existential rule classes which
could be non-destructive of arity-two QA.

\section{From Existential Rules to Arity-Two}
\label{sec:decide}
We will focus on the subclass of frontier-one rules whose heads
do not contain non-trivial \deft{Berge cycles} \cite{berge}.

\begin{definition}
  \label{def:berge}
  A \deft{Berge cycle} in a conjunction of atoms $\Psi$ is a sequence
  $A_1, x_1, \allowbreak A_2, x_2, \ldots, \allowbreak A_n, x_n$ of length $n > 1$ where the $x_i$ are pairwise distinct
  variables, the $A_i$ are pairwise distinct atoms of $\Psi$, and every $x_i$ occurs in
  atoms $A_i$ and $A_{i+1}$ (with addition modulo $n$, so $x_n$ occurs in $A_1$). 

  We say $\Psi$ is \deft{non-looping} if there is no Berge cycle of length above
  $2$, and no Berge cycle that contains an atom of~$\sigma_{>2}$.

  We define the \deft{head-non-looping} $\foneha$ subclass of $\fone$ rules
  whose heads are non-looping. In particular, single-head $\fone$ rules are
  always head-non-looping.
\end{definition}

\begin{example}
  Rules $A(x) \rightarrow \exists y z ~ R(x, y), S(y, z), T(z, x)$ and
  $B(y) \rightarrow \exists y z ~ R(x, y), U(x, y, z)$ are not in $\foneha$.
  However, $A(x) \rightarrow \exists y ~ V(x, x, y, y)$ and $B(x) \rightarrow
  \exists y ~ R(x, y), \allowbreak S(x, y), \allowbreak R(y, x)$ are in $\foneha$.
\end{example}

We claim that head-non-looping rules are non-destructive, in contrast with
general frontier-one rules (Theorem~\ref{thm:fronedl}):

\begin{theorem}
  \label{thm:foneha}
  $\foneha$ is not destructive of arity-two QA.
\end{theorem}

Of course, this means that QA is decidable for $\foneha \wedge \calD$, for any
DL $\calD$ expressible in $\gctwo$, such as $\alcqib$.
The rest of this section proves the theorem and addresses complexity.

\myparagraph{Shredding}
Our proof of Theorem~\ref{thm:foneha} translates the $\foneha$ rules to
arity-two constraints, using
a common way to represent general relational databases
in a binary relational store,
which  we call \deft{shredding}: we  represent
an $n$-ary relation by a set of binary relations giving the link
from each tuple (materialized as an element) to its attributes.
We present first the translation of the
signature~$\sigma$ to its shredded arity-two signature $\sigmas$, and the constraints imposed
on $\sigmas$-interpretations to ensure that they can
be decoded back to $\sigma$-interpretations. Second, we explain how to shred
facts and CQs.

\begin{definition}
The \deft{shredded signature} $\sigmas$ of a signature~$\sigma$
consists of $\sigma_{\leq 2}$, a unary relation $\elt$, and, for each
$R \in \sigma_{>2}$, a unary relation $A_R$ and binary relations
$R_i$ for $1 \leq i \leq \arity{R}$.

The \deft{well-formedness constraints} of $\sigmas$, written $\wf(\sigmas)$, are
the following DL constraints (they are $\alcqib$-expressible):
\begin{compactitem}
\item $C \sqsubseteq \elt$ for every unary relation $C$ of $\sigma_{\leq 2}$
\item $\exists R . \top \sqsubseteq \elt$ and $\exists R^- . \top \sqsubseteq
  \elt$ for all
  binary $R$ of~$\sigma_{\leq 2}$
\end{compactitem}
and the following, where $R \neq S$ are in $\sigma_{>2}$ and $1 \leq i
\leq \arity{R}$:
\begin{compactitem}
\item $\exists R_i . \top \sqsubseteq A_R$ and $\exists R_i^- . \top \sqsubseteq \elt$
\item $\elt \sqcap A_R \sqsubseteq \false$ and $A_R \sqcap A_S \sqsubseteq \false$
\item $A_R \sqsubseteq \exists R_i . \top$ and $\funct(R_i)$
\end{compactitem}
\end{definition}

The \deft{shredding} $\shred(F)$ of a $\sigma$-fact $F$ is the $\sigmas$-fact
obtained by adding the atom $\mathrm{Elt}(x)$ for each variable $x$ of $F$ and
replacing each atom $R(\mathbf{x})$ of $F$ when $R \in \sigma_{>2}$
by the atoms $A_R(t)$ and $R_i(t, x_i)$ for $1 \leq i \leq \arity{R}$, for
$t$ a fresh variable. The \deft{shredding}
$\shred(q)$ of a CQ $q$ is similarly defined.

\begin{example}
  Considering CQ $q: \exists x y z ~ U(x), \allowbreak R(x, y),  \allowbreak  S(z,
  z, x)$, we define $\shred(q)$ as:
  $\exists x y z t ~ \elt(x), \elt(y), \allowbreak \elt(z), U(x), \allowbreak
  R(x, y), \allowbreak A_R(t), \allowbreak S_1(t, z), \allowbreak S_2(t, z), \allowbreak S_3(t, x)$.
\end{example}

\myparagraph{Fully-non-looping}
The interesting part is to define the shredding of $\foneha$
rules. We first restrict to the class of \deft{fully-non-looping} rules,
$\fonehba$, whose head \emph{and body} are non-looping. We show that $\fonehba$
can be directly shredded to $\gctwo$.
We will later move from $\fonehba$ to $\foneha$.

For any existential
rule $\tau: \forall \mathbf{x} ~ \phi(\mathbf{x}) \rightarrow
\exists \mathbf{y} ~
\psi(\mathbf{x'}, \mathbf{y})$ with $\mathbf{x} \subseteq \mathbf{x'}$, we define its \deft{shredding}
$\shred(\tau)$ as the existential rule $\forall \mathbf{x} \mathbf{t} ~
(\shred(\phi(\mathbf{x}))) \rightarrow \exists \mathbf{y} \mathbf{t'} ~
(\shred(\psi(\mathbf{x'}, \mathbf{y})))$, where $\mathbf{t}$ and $\mathbf{t'}$
are the fresh elements introduced in the shredding of $\phi$ and $\psi$
respectively. We claim the following:

\begin{lemma}
  \label{lem:fonetogctwo}
  For any $\fonehba$ rule $\tau$, $\shred(\tau)$ can be translated in PTIME to a
  $\gctwo$ sentence on $\sigmas$.
\end{lemma}

\begin{example}
  For brevity, this example ignores the $\elt$ and $A_R$ atoms when shredding.
  Consider the $\fonehba$ rule:\\[.1cm]
  $U(u), T(u, x), S(x) \rightarrow \exists y z ~ T(x,
  y), U(y), R(x, x, z, z)$\\[.1cm]
  Its shredding is expressible in $\gctwo$ (and even in $\alcqib$):\\[.1cm]
  $(\exists T^- . U) \sqcap S \sqsubseteq (\exists T . U) \sqcap
  (\exists (R_1^- \sqcap R_2^-) . (\exists (R_3 \sqcap R_4) . \top))$\\[-.3cm]

    By contrast, consider the following rule in $\fone \backslash
    \foneha$:\\[.1cm]
    $U(x) \rightarrow \exists y z ~ R(x, y), S(x, y, z)$\\[.1cm]
    Its shredding is as follows; it is not $\gctwo$-expressible:\\[.1cm]
    $U(x) \rightarrow \exists y z t ~ R(x, y), S_1(t, x), S_2(t, y), S_3(t,
    z)$
\end{example}

In the general case, the $\gctwo$ rewriting of Lemma~\ref{lem:fonetogctwo} is obtained in PTIME by seeing
the body and head of $\shred(\tau)$ as a tree, which is possible because
$\tau$ is fully-non-looping.

It is now easy to show the following general result:

\begin{proposition}[Shredding]
  \label{prp:fonehba}
For any
fact~$F$, $\gctwo$ constraints
$\Sigma$, existential rules $\Delta$ and CQ $q$, the following are
equivalent:
\begin{compactitem}
\item $F \wedge \Sigma \wedge \Delta \entunr q$;
\item $\shred(F) \wedge \Sigma \wedge \shred(\Delta) \wedge
  \wf(\sigmas) \entunr \shred(q)$.
\end{compactitem}
\end{proposition}

Thus, from Lemma~\ref{lem:fonetogctwo}, as
$\shred(F)$, $\shred(\Delta)$, $\sigmas$, $\wf(\sigmas)$, and $\shred(q)$ can be computed
in PTIME following their definition, we deduce the following,
in the case of $\fonehba$:

\begin{corollary}
  QA for $\gctwo$ and $\fonehba$ constraints can be reduced to QA for $\gctwo$
  in PTIME; further, when the constraints and query are fixed in the input,
  they also are in the output, so data complexity bounds for $\gctwo$ QA
  are preserved.
\end{corollary}

This concludes the proof of Theorem~\ref{thm:foneha} for $\fonehba$
constraints. It further implies that QA for $\gctwo$ and $\fonehba$ has co-NP-complete data
complexity, like $\gctwo$, \cite{pratt2009data}, and the combined complexity is
the same as for $\gctwo$.

Note that, although QA for $\gctwo$ is decidable, we know of no  realistic implementations.
Our translation could however reduce instead to arity-two QA with constraints in
DLs such as $\alcqib$, if we impose
impose  additional minor restrictions on the $\fonehba$ rules (e.g., no atom of the form $S(x, x)$).
For simplicity, however, we focus in the sequel on reductions to
decidable QA on arity-two (i.e., translating to $\gctwo$) rather than
investigating which restrictions would ensure that the output of
our translations can be expressed in particular DLs.

\myparagraph{Head-non-looping}
We now extend the claim to $\foneha$ rather than $\fonehba$.
The idea is that we
rewrite $\foneha$ rules to $\fonehba$ by \deft{treeifying them}, considering all
possible fully-non-looping rules that they imply, and all possible ways that
they
can match on the parts of the interpretations that satisfy the fact.
To formalize this, we assume that
we have added to the fact $F$ one atom $P_x(x)$ for each variable $x$ of $F$, where each $P_x$ is a
fresh unary relation.
We then define:

\begin{definition}
  \label{def:treeif}
  The \deft{treeification} on fact~$F$ of a $\foneha$ rule $\tau: \forall
  \mathbf{x} ~ (\phi(\mathbf{x}) \rightarrow \exists \mathbf{y} ~
  \psi(x_{\f}, \mathbf{y}))$, where $x_{\f} \in \mathbf{x}$ is the frontier
  variable, is the
conjunction $\treeif_F(\tau)$ of $\fonehba$ rules defined as follows:
\begin{compactitem}
  \item consider every mapping $f$ from $\mathbf{x}$ to
    itself, and let $f(\tau)$ be obtained from $\tau$ by renaming all
    variables in $\mathbf{x}$ with~$f$;
  \item for every such $f(\tau)$, consider every $\mathbf{x'} \subseteq
    \mathbf{x}$
    and every mapping $g$ from $\mathbf{x'}$ to the variables of~$F$,
    and construct $g(f(\tau))$
    by replacing every occurrence of each $x \in
    \mathbf{x'}$ in $\phi(\mathbf{x})$ by fresh variables $x_1, \ldots, x_n$, and
    adding the facts $P_{\smash{g(x)}}(x_i)$ for all $x \in
    \mathbf{x'}$ and all~$i$ (if $x_{\f} \in \mathbf{x'}$, also replace $x_{\f}$ in
    $\psi(x_{\f}, \mathbf{y})$ by one of its copies);
  \item if $g(f(\tau))$ is fully-non-looping, add it to
    $\treeif_F(\tau)$.
\end{compactitem}
\end{definition}

\begin{example}
  Consider a fact~$F$ and the following rule~$\tau$:\\[0.1cm]
  $R(x, y), S(y, z), T(z, w),
  U(w, x) \rightarrow A(x)$\\[0.1cm]
  The treeification $\treeif_F(\tau)$ contains the rule:\\[0.1cm]
  $R(x, y), S(y, z), T(z, y), U(y, x)
  \rightarrow A(x)$.
  \smallskip

  Consider the rule $\tau': R(x, y), S(y, x, x) \rightarrow A(x)$,
  and a fact~$F$ containing variable~$z$.
  Then $\treeif_F(\tau')$ contains:\\[0.1cm]
  $R(x_1, y), S(y, x_2, x_3), P_z(x_1), P_z(x_2),
  P_z(x_3) 
  \rightarrow A(x_1)$
\end{example}

We now claim:

\begin{proposition} \label{prp:treeify}
For any 
fact $F$, $\gctwo$ constraints
$\Sigma$, $\foneha$ rules $\Delta$ and CQ $q$, the following are
equivalent:
\begin{compactitem}
\item $F \wedge \Sigma \wedge \Delta \entunr q$;
\item $F \wedge \Sigma \wedge \treeif_F(\Delta) \entunr q$.
\end{compactitem}
\end{proposition}

This proposition implies that QA for $\foneha$ and $\gctwo$ can be reduced to QA
for $\fonehba$ and $\gctwo$, which is decidable by the Shredding Proposition,
proving Theorem~\ref{thm:foneha}.

To prove Proposition~\ref{prp:treeify}, for the first direction, if
$F \wedge \Sigma \wedge \Delta \not\entunr q$, one can show that all of the
fresh unary relations~$P_x$ in an interpretation of $F \wedge \Sigma \wedge
\Delta \wedge \neg q$ can be assumed to be interpreted by one tuple. One then
shows that
$\Delta$ implies $\treeif_F(\Delta)$ on such interpretations.
For the other direction, assuming that
$F \wedge \Sigma \wedge \treeif_F(\Delta) \not\entunr q$,
the Shredding
Proposition implies that there is
 a $\sigmas$-interpretation $\calJ$ of
$\Theta \defeq \Sigma \wedge \shred(\treeif_F(\Delta)) \wedge \wf(\sigmas)$,
$\neg q' \defeq \neg \shred(q)$, and the existential closure of $F' \defeq \shred(F)$. 
We 
apply an unraveling argument
to show that $\calJ$ can be made \deft{cycle-free}:

\begin{definition}
  \label{def:cyclef}
  The \deft{Gaifman graph} $\gaifman{\calI}$ of an interpretation $\calI$ is the undirected graph on
  $\dom(\calI)$ connecting any two elements co-occurring in a tuple of
  $\calI$. Given a fact $F$, an interpretation $\calI$ is
  \deft{cycle-free except for $F$} if $F$ has a witness~$\calW$ in~$\calI$
  such that any cycle of~$\gaifman{\calI}$ is only on elements
  of~$\dom(\calW)$.
\end{definition}

\begin{lemma}[Unraveling]
  \label{lem:unravel}
  For any $\sigmas$-fact $F'$, $\gctwo$ constraints $\Theta$, and CQ
  $q'$,
  if $(\exists \mathbf{x} \mathbf{t} ~ F'(\mathbf{x}, \mathbf{t})) \wedge \Theta \wedge \neg q'$ is satisfiable then it has an
  interpretation which is cycle-free except for $F'$.
\end{lemma}

Letting $\calJ'$ be the
\deft{unraveling} of our interpretation~$\calJ$ (obtained by the Unraveling
Lemma), we can then ``unshred'' $\calJ'$
back to a $\sigma$-interpretation $\calI$:

\begin{definition}
  \label{def:unshredint}
  The \deft{unshredding} $\calI$ of a $\sigmas$-interpretation $\calJ \models
  \wf(\sigmas)$ is obtained by setting $R^{\calI} \defeq R^{\calJ}$ for $R \in
  \sigma_{\leq 2}$, and, for all $R \in \sigma_{>2}$ and $t \in A_R^{\calJ}$,
  creating the tuple $\mathbf{a} \in R^{\calI}$ such that $(t, a_i) \in R_i^{\calJ}$
  for all $1 \leq i \leq \arity{R}$.
\end{definition}

As in the proof of the Shredding Proposition, we can show that the unshredding $\calI$
is well-defined and satisfies the unshredded constraints $(\exists \mathbf{x} ~
F(\mathbf{x})) \wedge \Sigma \wedge
\treeif_F(\Delta) \wedge \neg q$. Further, we show that it satisfies
$\Delta$ and not just $\treeif_F(\Delta)$, because
a match of a $\foneha$ rule $\tau$ in $\calI$ must be a match of
$\treeif_F(\tau)$; otherwise the match would witness that $\calJ'$ was not cycle-free:

\begin{lemma}[Soundness]
  \label{lem:preserv2}
  For a $\sigma$-fact $F$, $\foneha$ rule~$\tau$ and $\sigmas$-interpretation
  $\calJ$, if $\calJ$ satisfies $\shred(\treeif_F(\tau))$ and is cycle-free except for
  $\shred(F)$, then the unshredding $\calI$ 
of~$\calJ$ satisfies $\tau$.
\end{lemma}

We conclude by sketching the proof of the Unraveling Lemma, which follows 
\cite{kazakov2004polynomial,pratt2009data}. From an interpretation $\calJ$ of
$(\exists \mathbf{x} \mathbf{t} ~ F'(\mathbf{x}, \mathbf{t})) \wedge \Theta \wedge \neg q'$, for all $u \neq v$ in
$\dom(\calJ)$ co-occurring in some tuple of~$\calJ$, we call a \deft{bag} 
the interpretation with domain $\{u, v\}$ consisting of the tuples of
$\calJ$ mentioning only $u, v$.
We build a graph $G$ over the
bags by connecting bags whose domain shares one element.
We pick a witness $\calW$ of~$F'$ in~$\calJ$ and
merge in the \deft{fact bag} all bags whose domain is included in $\dom(\calW)$.

An unraveling is a tree $T$ of bags obtained by unfolding $G$
starting at the fact bag, which is preserved as-is. Each bag $b$ of~$T$ except the fact bag has a
domain containing two elements: one of them occurs exactly in $b$, its siblings
and its parent; the other occurs exactly in $b$ and its
children (it is \deft{introduced} in~$b$). We see $T$ as an interpretation formed of
the union of its bags.

We construct $T$ from $G$ inductively. For any bag $b$ in $T$ corresponding
to a bag $b'$ in $G$, construct the children of~$b$ as follows. For each bag
$b''$ adjacent to $b'$ in $G$, if $b'$ and $b''$ share the element corresponding
to the element $u$ introduced in $b$, create an isomorphic copy of~$b''$ as a child of~$b$ in $T$, whose domain
is $u$ plus a fresh element, and perform the
unraveling process recursively on the children.

It can be shown that the unraveling operation preserves $\gctwo$ constraints,
the fact~$F'$, and the negated CQ $\neg q'$.
As $T$ is a tree, the interpretation it describes is
cycle-free (except for the witness $\calW$, because we copied the fact bag
as-is).

\myparagraph{Complexity}
Proposition~\ref{prp:treeify} gives a reduction from $\foneha$ and $\gctwo$ QA
to $\fonehba$ to $\gctwo$ QA, but its
output is of exponential size in the input, because of treeification. Hence, letting $\mathfrak{f}(n)$
bound the size of the output of our reduction
given an input of size $n$, and letting
$\mathfrak{g}(n)$ bound the combined complexity of
$\gctwo$ QA, we have shown an upper bound
of $\mathfrak{g}(\mathfrak{f}(n))$ for QA for $\foneha$ and $\gctwo$.

Further, treeification rewrites the rules in a fact-dependent
way, so, unlike the previous case of $\fonehba$ and
$\gctwo$ QA, data complexity bounds for $\gctwo$ QA do not imply
data complexity bounds for $\foneha$ and $\gctwo$ QA.

\section{Adding Functional Dependencies}
\label{sec:fds}
The previous section showed that the language of head-non-looping
frontier-one rules is not destructive of $\gctwo$ QA. However, another kind of
rules that we would want to support on higher-arity relations are
functional dependencies (FDs).

It is well-known that QA is undecidable for, e.g.,
$\bidclass$ and 
arbitrary FDs
\cite{cali2003decidability}, so such constraints are trivially destructive. As it turns
out, undecidability also holds for $\foneha$ rules and FDs; in fact, even for
single-head $\foneldclass$ rules and FDs:

\begin{theorem}
  \label{thm:fdsfrone}
  QA is undecidable for FDs and single-head frontier-one rules, even if all FDs
  have a determiner of size~$1$.
\end{theorem}

However, for certain kinds of existential rules and FDs, QA is known to be
decidable: this is in particular the case of \deft{non-conflicting} rules and
FDs \cite{cali2012towards}:

\begin{definition}
  We say that a \emph{single-head} existential rule $\tau$ is \emph{non-conflicting} with
respect to a set of FDs
$\Phi$ if, letting $A =
R(\mathbf{z})$ be the head atom of~$\tau$, letting $S$ be the subset of $\{1, \ldots,
\arity{R}\}$ such that $z_i$ is a frontier variable iff $i \in S$:
\begin{compactitem}
\item No strict subset of $S$ is the determiner of an FD in $\Phi$;
\item If $S$ is exactly the determiner of an FD of $\Phi$, then all existentially
quantified variables in $A$ occur only once.
\end{compactitem}
\end{definition}

Note that this requires rules to be \emph{single-head}, and thus
head-non-looping. 
Our result with respect to adding FDs is:

\begin{theorem}
  \label{thm:noconflict}
Non-conflicting frontier-one rules and FDs are non-destructive of
arity-two QA.
\end{theorem}

In particular, single-head frontier-one rules and FDs are non-destructive of
arity-two QA if all variables in the head atom of rules are assumed to have only one
occurrence, as this simple sufficient condition implies the
non-conflicting condition.

To prove the theorem,  we assume without loss of generality that we only
have FDs on higher-arity relations, as we can write them in $\gctwo$ otherwise.
We cannot shred the FDs, as they would translate to a
functionality assertion for the path, e.g., $R_i^- \circ R_j$, which is not expressible
in $\gctwo$ (and not even in expressive DLs such as
$\sroiq$~\cite{horrocks2006even}).
However, we can show that, thanks to the non-conflicting requirement, FDs
can always be made to hold on interpretations, as long as they hold on a witness
of the fact.

\begin{proposition}
  \label{prp:addfds}
  For any $\gctwo$ constraints $\Sigma$, non-conflicting
  frontier-one rules $\Delta$, FDs $\Phi$ on~$\sigma_{>2}$, $\sigma$-fact $F$,
  and CQ $q$, if there is an interpretation $\calI$ satisfying $\Theta \defeq
  (\exists \mathbf{x} ~ F(\mathbf{x})) \wedge \Sigma \wedge \Delta 
  \wedge \neg q$ and  there is a witness $\calW$ of $F$ in $\calI$ satisfying $\Phi$,
  then $\Theta \wedge \Phi$ is satisfiable.
\end{proposition}

We first prove Proposition~\ref{prp:addfds}.
As in Section~\ref{sec:decide}, consider the treeification $\treeif_F(\Delta)$: it is still
non-conflicting as treeification only affects rule bodies.
Use the Shredding Proposition to obtain an interpretation $\calJ$ of
$\neg q' \defeq \neg \shred(q)$,
$\Theta \defeq \Sigma \wedge \shred(\treeif_F(\Delta)) \wedge \wf(\sigmas)$,
and the existential closure of $F' \defeq
\shred(F)$.
By our
hypothesis about the existence of a witness, we can assume that $\calJ$
has a witness $\calW$ of $F'$ whose unshredding satisfies~$\Phi$.

In the previous
section, we used the Unraveling Lemma to show that $\calJ$ could be assumed
to be cycle-free. We now modify the lemma to additionally ensure the following
property on~$\calJ$, which will forbid FD violations in its unshredding:

\begin{definition}
  Given a set of FDs $\Phi$ on~$\sigma_{>2}$, a
  $\sigmas$-interpretation $\calJ$, and a witness $\calW$ of a fact in~$\calJ$, we call $\calJ$
  \deft{FD-safe except for $\calW$} if for every $a \in \dom(\calJ)$, 
  for any $R \in \sigma_{>2}$ and FD determiner $P$ of $R$ in
  $\Phi$, considering each $t \in \dom(\calJ)$ such that 
  $(t, a) \in R_i^{\calJ}$ for every $i \in P$, either there is at most one
  such~$t$ or all are in $\dom(\calW)$.
\end{definition}

FD-safety is useful for the following reason:

\begin{lemma}
  \label{lem:whysafe2}
  For any set of FDs $\Phi$ on~$\sigma_{>2}$, for any $\sigmas$-interpretation $\calJ$
  which is cycle-free and FD-safe except for a witness $\calW$, if the unshredding of
  $\calW$ satisfies $\Phi$, then the unshredding of $\calJ$ satisfies $\Phi$.
\end{lemma}

We now claim a variant of the Unraveling Lemma:

\begin{lemma}[FD-aware unraveling]
  \label{lem:unravelb}
  Let $\Sigma$ be a $\gctwo$ constraint, $F$ a $\sigma$-fact,
  $q$ a CQ, $\Delta$ non-conflicting
  frontier-one rules and $\Phi$ a set of FDs on~$\sigma_{>2}$. Let 
  $\calJ$ be an interpretation satisfying  $\Theta \defeq
  (\exists \mathbf{x} \mathbf{t} ~ \shred(F)(\mathbf{x}, \mathbf{t}))
  \wedge \Sigma \wedge
  \shred(\treeif_F(\Delta)) \wedge \wf(\sigmas) \wedge \neg \shred(q)$, and
  $\calW$ a witness of $\shred(F)$ in $\calJ$.
  Then there is an interpretation $\calJ'$ satisfying $\Theta$ such that $\calW$
  is a witness
  of~$\shred(F)$ in~$\calJ'$, and $\calJ'$ is cycle-free and FD-safe except for
  $\calW$.
\end{lemma}

We prove the lemma by tweaking the unraveling process to ensure FD-safety: when
creating children of each bag~$b$ in the unraveling $T$ for
neighbors of its corresponding bag~$b'$ in the bag graph $G$, omit some
neighbors that contain shreddings of higher-arity tuples if
the shared element~$u$ occurs in a strict superset of an FD determiner
of~$\Phi$, and unravel differently the neighbors where $u$ occurs
exactly at a determiner.
This unraveling still satisfies
$\Sigma$, $\neg q'$, and the existential closure of~$F'$, and satisfies $\shred(\treeif_F(\Delta))$: the
non-conflicting condition ensures that the omitted facts were not required by a
rule.

We then apply the FD-aware Unraveling Lemma to $\calJ$ and consider the
unshredding
$\calI$ of the result; it satisfies all necessary constraints as in
Section~\ref{sec:decide}, including $\Phi$ by Lemma~\ref{lem:whysafe2}. This proves Proposition~\ref{prp:addfds}.

We conclude by proving Theorem~\ref{thm:noconflict}. We first observe that the
results of Section~\ref{sec:decide} extend to a more general notion of fact
that allows inequality axioms ($x \neq y$); indeed, inequalities in the fact are preserved
by shredding and unshredding, and by unraveling.
So Theorem~\ref{thm:foneha} holds for such facts with
inequalities, with the same complexity. Second, we enumerate all possible
equalities between variables of the fact~$F$, and for each possibility, consider
the fact~$F_=$ where variables are merged following the equalities, and
inequalities are asserted between the remaining variables. 
Proposition~\ref{prp:addfds} implies that our original entailment holds iff
all the derived entailments hold where $F$ is replaced by some $F_=$ whose canonical
interpretation satisfies~$\Phi$ (this can be tested in PTIME for each~$F_=$). Thus we have reduced to QA for $\fonehba$ and $\gctwo$.

In terms of complexity, as $\gctwo$ QA is EXPTIME-hard in combined complexity
(because satisfiability for the usual two-variable
guarded fragment is EXPTIME-hard \cite{gradel1999restraining}), the additional exponential
factor (from all possible $F_=$) 
has no impact, so the bounds of Section~\ref{sec:decide} also apply to QA for
$\gctwo$ and non-conflicting frontier-one rules and FDs.

\section{Conclusion}
\label{sec:conc}

In this paper, we have studied the impact
of existential rules on the decidability of query answering
for classes of arity-two constraints.
We also explained (in proving Theorem \ref{thm:noconflict}) how the decidability extends
when inequalities are allowed in  facts.

We have limited our arbitrary arity constraints 
to rules, i.e., dependencies. In future work we will
study how to extend our 
results to arbitrary arity constraint languages 
with more features, e.g., disjunction.  We will also study what happens
in the presence of
constants (or nominals), which are disallowed in $\gctwo$ (and in the rule languages we consider), 
but are known not to break decidability in arity-two
contexts \cite{nominalsinverses,nominalspath}. This, however, would probably
require different techniques, as unraveling may create multiple copies of
constants. 
Another question that would probably require specific tools is
the study of \emph{finite QA}, i.e., QA restricted to finite interpretations.

\paragraph{Acknowledgements.}
We are very grateful to Boris Motik and Pierre
Senellart for their helpful feedback. This work was partly supported by
the T\'el\'ecom ParisTech Research Chair
on Big Data and Market Insights and by
the Engineering and Physical Sciences Research Council, UK, grants EP/G004021/1
and EP/M005852/1.

\bibliographystyle{alpha}
\bibliography{unr}

\appendix
\makeatletter
\newenvironment{proofb}[1][\proofname]{\par
  \vspace{-\topsep}
  \pushQED{\qed}%
  \normalfont
  \topsep2pt \partopsep1pt 
  \trivlist
  \item[\hskip\labelsep
        \itshape
    #1\@addpunct{.}]\ignorespaces
}{%
  \popQED\endtrivlist\@endpefalse
  \addvspace{6pt plus 6pt} 
}
\makeatother
\section{Proofs for Section~\ref{sec:undecide}: Negative Results for Combination}

\subsection{Proof of Theorem~\ref{thm:wadl}: $\stt$ is destructive}

Adapt the proof of Theorem~\ref{thm:fronedl} by rewriting $\tau$ to replace all
$S$-atoms in the right-hand side by $S'$-atoms. The resulting rule is
clearly source-to-target, with $\sigma_\S = \{S\}$ and $\sigma_\T = \{D, R,
S'\}$. Now impose the concept inclusion $S' \sqsubseteq S$. It is clear that the
resulting rules are equivalent to those of Theorem~\ref{thm:fronedl}, so the
same proof applies.

\subsection{Proof of Theorem~\ref{thm:biddl}: $\bidclass$ is destructive}
  \emph{In this section, as in the rest of the appendix, we write the positions of
any relation $R$ as $R^1, \ldots, R^{\arity{R}}$.}
\medskip

We will show this undecidability result by considering the \deft{entailment
problem}.

\begin{definition}
  \label{def:ent}
  The \deft{(unrestricted) entailment problem} for two classes $\class_1,
  \class_2$ of
  constraints, asks, given a set of constraints $\Sigma$ of $\class_1$ and a
  constraint $\tau \in \class_2$, whether $\Sigma$ \deft{entails} $\tau$, written
  $\Sigma \entunr \tau$. That is, whether any interpretation of $\Sigma$ is an
  interpretation of
  $\tau$.
\end{definition}

We show a reduction to QA for a class of logical constraints to entailment to this class
of constraints and rules. The idea follows \cite{cali2003query}
(Theorem~3.4) but is slightly more complicated to take care of a difficulty that
was omitted there.

\begin{lemma}
  \label{lem:qatoimp}
  For any class $\class_1$ of constraints and $\class_2$ of existential rules,
  there is a reduction from entailment for $\class_1$ and $\class_2$ to QA for
  $\class_1$.
\end{lemma}

\begin{proofb}
  Consider an instance of the entailment problem for $\class_1$ and $\class_2$: $\Sigma$ is
  a set of constraints of $\class_1$, and $\tau : \forall \mathbf{x} ~
  (\phi(\mathbf{x}) \rightarrow \exists \mathbf{y} ~ \psi(\mathbf{x'},
  \mathbf{y}))$ with $\mathbf{x'} \subseteq \mathbf{x}$ is an existential rule of
  $\class_2$. Let us reduce this to an instance of the QA problem for $\class_1$.

  Create fresh unary relations $P_x$ for each $x \in \mathbf{x}$.
  We consider the QA instance asking whether $F \wedge \Sigma \entunr q$,
  where the fact~$F$ is $\phi(\mathbf{x}) \wedge \bigwedge_{x \in \mathbf{x}} P_x(x)$
  and the query~$q$ is $\exists \mathbf{x} \mathbf{y} ~
  \phi(\mathbf{x}) \wedge \psi(\mathbf{x}', \mathbf{y}) \wedge \bigwedge_{x \in
  \mathbf{x}} P_{x}(x)$. We claim that $F \wedge \Sigma \entunr
  q$ iff $\Sigma \entunr \tau$, which proves that the reduction is correct.
  
  If $\Sigma \entunr \tau$, then consider an
  interpretation $\calI$ satisfying $\Sigma$ and the existential closure of~$F$.
  As $\calI \models \Sigma$ and
  $\Sigma \entunr \tau$, we have $\calI \models \tau$; thus, applying $\tau$ to
  any witness of~$F$ in~$\calI$, we
  deduce the existence of a match of $q$. This proves that $F \wedge \Sigma \entunr q$.
  
  Conversely, if $\Sigma \not\entunr \tau$, there exists an interpretation
  $\calI$ of
  $\Sigma$ that does not satisfy $\tau$, meaning that there is a violation of
  $\tau$ in $\calI$: a set $\mathbf{b}$ of elements of $\dom(\calI)$ such that
  $\calI \models \phi(\mathbf{b})$ but this match cannot be extended to a match of
  $\psi$. Let us modify $\calI$ to $\calI'$ by setting, for each
  $x \in \mathbf{x}$,
  $P_x^{\calI'} \defeq \{(b)\}$, where $b$ is the element of
  $\mathbf{b}$ corresponding to $x \in \mathbf{x}$, and setting $R^{\calI'}
  \defeq R^{\calI}$ for all other relations $R$. It is clear that $\calI'$ 
  still satisfies
  $\Sigma$, as $\Sigma$ does not mention the fresh unary relations $P_x$. Now,
  we also have $\calI' \models \phi(\mathbf{b})$, and by construction $\calI'
  \models \bigwedge_{b \in \mathbf{b}} P_x(b)$, so that $\calI'$ satisfies the
  existential closure of $F$. However, $\calI'$ does not satisfy $q$:
  the only possible
  match of $q$ is on the elements that occur in the $P_x^{\calI'}$, and the impossibility to extend this
  match to a match of $\psi$ is by definition of it being a violation of $\tau$.
  Hence, $\calI'$ witnesses that $F \wedge \Sigma \not\entunr \tau$.

  Hence, the reduction is correct, which concludes the proof.
\end{proofb}

Thus, let $\calD$ be a DL that can express the assertions $\funct(R)$
for any binary relation~$R$. To show the undecidability of QA for
$\calD \wedge \bidclass$, by the above, it suffices to show the undecidability
of entailment for $\calD \wedge \bidclass$ and $\bidclass$.

\begin{definition}
  \label{def:ufd}
We call $\ufdclass$ the class of \deft{unary functional dependencies} (UFDs),
that is, functional dependencies (on arbitrary arity relations) whose
determiner consist of a single attribute. We write UFDs as $R^p \rightarrow
R^q$, where $R^p$ and $R^q$ are positions of a higher-arity relation $R$.
\end{definition}

We now claim that functionality assertions on binary relations can be
bootstrapped to UFDs on arbitrary arity relations, using $\bidclass$:

\begin{lemma}
  \label{lem:liftfd}
  There is a reduction from entailment for $\ufdclass \wedge \bidclass$ and
  $\bidclass$ to entailment for $\calD \wedge \bidclass$ and $\bidclass$.
\end{lemma}

\begin{proofb}
  Consider constraints $\Sigma$ of $\ufdclass \wedge \bidclass$
  and a rule $\tau \in \bidclass$.
  Encode each UFD $\phi: R^p \rightarrow R^q$ of $\Sigma$
  as an $\bidclass$ rule $\tau_\phi: \forall
  \mathbf{x} ~ R(\mathbf{x}) \rightarrow R_{\phi}(x_p, x_q)$,
  where $R_{\phi}$ is a fresh binary relation for $\phi$, and a functionality
  assertion $\funct(R_{\phi})$. Let the constraints $\Sigma'$ consist of the original
  $\bidclass$ rules, the new $\bidclass$ rules, and the
  functionality assertions. We claim that $\Sigma \entunr \tau$ iff
  $\Sigma' \entunr \tau$.

  If $\Sigma' \not\entunr \tau$, let $\calI$ be a counterexample interpretation satisfying
  $\Sigma'$ but not~$\tau$. We claim that $\calI$ also satisfies $\Sigma$. Indeed,
  the only thing to check is that UFDs are satisfied; but assume that there is a
  UFD $\phi:R^p \rightarrow R^q$ of $\Sigma$ that has a violation in $\calI$,
  namely, two tuples $\mathbf{a}, \mathbf{b} \in R^{\calI}$ such that $a_p = b_p$
  but $a_q \neq b_q$. As $\calI$ satisfies the $\bidclass$ rule $\tau_\phi$, we have
  $(a_p, a_q) \in R_\phi^{\calI}$ and $(b_p, b_q) \in R_{\phi}^{\calI}$;
  this contradicts the
  assertion $\funct(R_\phi)$ that $\calI$ is supposed to respect. Hence $\calI$
  satisfies $\Sigma$, and as $\calI$ does not satisfy $\tau$, it witnesses that
  $\Sigma \not\entunr \tau$.

  Conversely, if $\Sigma \not\entunr \tau$, let $\calI$ be a counterexample
  interpretation
  satisfying $\Sigma$ but not~$\tau$. Without loss of generality, we have
  $R_{\phi}^{\calI} = \emptyset$ for all the fresh relations $R_{\phi}$, as they
  are not mentioned in $\Sigma$. Now, extend $\calI$ to an interpretation $\calI'$
  that satisfies $\Sigma'$ by adding to $R_\phi^{\calI'}$, for every FD $\phi: R^p \rightarrow R^q$, for every
  $\mathbf{a} \in R^{\calI}$, the tuple $(a_p, a_q)$. It is clear that the
  result $\calI'$ still satisfies the $\bidclass$ rules of $\Sigma$, and that it
  satisfies the $\bidclass$ rules of $\Sigma'$; and it is easily seen
  that it satisfies the functionality assertions as otherwise, as before, a
  violation of such an assertion in~$\calI'$ witnesses a violation of the UFDs
  of $\Sigma$ in~$\calI$.
  Further, as $\tau$ does not mention the $R_\phi$, $\calI'$ still does not satisfy
  $\tau$, because $\calI$ did not. Hence, $\calI'$ witnesses that $\Sigma' \not\entunr \tau$.

  This shows that the reduction is correct, concluding the proof.
\end{proofb}

\begin{definition}
  \label{def:uid}
  The class of \deft{frontier-one inclusion dependencies} (or \emph{unary}
  inclusion dependencies), $\uidclass$, is the class of inclusion dependencies
  with frontier of size~$1$.
  We write an $\uidclass$ rule $\forall \mathbf{x} ~ (R(\mathbf{x})
  \rightarrow \exists \mathbf{y} ~ S(\mathbf{x}', \mathbf{y}))$ as $R^p
  \subseteq S^q$, where $R^p$ and $S^q$ are the positions at which the
  frontier variable occurs in the body and head atom respectively.

  Following this convention, we write rules of $\bidclass$ in the same
  way: $R^a R^b \subseteq S^c S^d$ denotes the rule $\forall \mathbf{x} ~ (R(\mathbf{x})
  \rightarrow \exists \mathbf{y} ~ S(\mathbf{x}', \mathbf{y}))$ where the
  first frontier variable occurs at positions $R^a$ and $S^c$ in the body and
  head, and the second occurs at positions $R^b$ and $S^d$ in the body and head.
  (Remember that the definition of $\idclass$ requires each variable to only
  occur once in the body atom and head atom.)
  Note that we must have $R^a \neq R^b$ and $S^c \neq S^d$; but we may have $R^a
  = S^c$ or $R^a = S^d$, and similarly for $R^b$.
\end{definition}

We now explain that we can add without loss of generality \emph{frontier-one
inclusion dependencies} (or \emph{unary} inclusion dependencies), $\uidclass$,
to the entailment problem, the reason being that $\uidclass$ rules can
be encoded in $\bidclass$ up to adding additional attributes.

\begin{lemma}
  \label{lem:rmuid}
  There is a reduction from entailment for $\ufdclass \wedge \uidclass \wedge \bidclass$ and
  $\bidclass$ to entailment for $\ufdclass \wedge \bidclass$ and $\bidclass$.
\end{lemma}

\begin{proofb}
  Consider constraints $\Sigma$ of $\ufdclass \wedge \uidclass \wedge
  \bidclass$ and $\tau \in \bidclass$. Let $\sigma_+$ be the signature obtained
  from $\sigma$ in the following way: for each relation $R \in \sigma$, we
  create a relation $R_+$ in $\sigma_+$ whose positions are those of $R$ plus
  one position $R_+^{\delta,1}$ for each $\uidclass$ rule~$\delta$ of the form $R^p \subseteq
  S^q$, and one position $R_+^{\delta,2}$ for each $\uidclass$ rule~$\delta$ of the form $S^q
  \subseteq R^p$.
  
  Now, encode each $\uidclass$ rule $\delta: R^p \subseteq
  S^q$ of $\Sigma$ as the following $\bidclass$ rule on $\sigma_+$: $R_+^p
  R_+^{\delta,1} \subseteq S_+^q S_+^{\delta,2}$.
  We thus define the constraints $\Sigma'$ on $\sigma_+$ to
  consist of these additional $\bidclass$ rules, and of the straightforward
  rewriting of the original $\bidclass$ and $\ufdclass$ constraints of $\sigma$
  to $\sigma_+$, rewriting, e.g., $R^a R^b \subseteq S^c S^d$ as $R_+^a R_+^b
  \subseteq S_+^c S_+^d$, and $R^p \rightarrow R^q$ as $R^p_+ \rightarrow
  S^q_+$.
  Once again we show that $\Sigma \entunr \tau$ iff $\Sigma' \entunr \tau$.

  If $\Sigma \not\entunr \tau$, then we extend a counterexample $\sigma$-interpretation
  $\calI$ to a $\sigma_+$-interpretation $\calI'$ satisfying $\Sigma'$ as
  follows:
  for all $R \in \sigma$, consider each tuple $\mathbf{a} \in R^{\calI}$,
  and create in $R_+^{\calI'}$ the tuple $\mathbf{b}$ defined by $b_p \defeq
  a_p$ for all positions $R^p$ of $R$, and $b_{\delta,1} \defeq a_p$ such that
  $\delta$ is of the form $R^p \subseteq S^q$, and $b_{\delta,2} \defeq a_p$ such
  that $\delta$ is of the form $S^q \subseteq R^p$.
  It is clear that the result
  $\calI'$ still satisfies the
  $\ufdclass$ and $\bidclass$ constraints of $\Sigma'$ and violates $\tau$,
  because they do not mention the new attributes of $\sigma_+$. Further,
  $\calI'$ clearly satisfies the new $\bidclass$ rules because the original
  interpretation $\calI$ satisfied the $\uidclass$ rules. Hence, $\calI'$ witnesses that
  $\Sigma' \not\entunr \tau$.

  Conversely, if $\Sigma' \not\entunr \tau$, we rewrite a counterexample
  $\sigma_+$-interpretation to a $\sigma$-interpretation of~$\Sigma$ by simply
  removing the additional attributes in all tuples, which
  clearly gives an interpretation satisfying $\Sigma$: it satisfies the
  $\uidclass$ rules because $\calI'$ satisfied the new $\bidclass$ rules of
  $\Sigma$, and the other constraints are preserved. This concludes the
  proof.
\end{proofb}

We are now ready to conclude, because:

\begin{theorem}[\cite{mitchell1983implication}]
  \label{thm:mitchell}
  The entailment problem for $\ufdclass \wedge \uidclass \wedge \bidclass$ and
  $\bidclass$ is undecidable.
\end{theorem}

This is a slightly stronger result than what is claimed in
\cite{mitchell1983implication}, because their definition of $\bidclass$
does not forbid repetitions of positions (i.e., it allows $\bidclass$ rules of the
form $R^p R^q \subseteq R^r R^r$). We refer to
Appendix~\ref{apx:prf_fdsfrone} for more details about how the stronger result
is proved.

This concludes the proof of Theorem~\ref{thm:biddl}, because, if QA for
$\bidclass \wedge \calD$ were decidable, then we would have decidability of the
entailment problem above, by reducing it successively through
Lemma~\ref{lem:rmuid}, Lemma~\ref{lem:liftfd}, and Lemma~\ref{lem:qatoimp}.

\subsection{Proof of Theorem~\ref{thm:fronedl}: $\fone$ is destructive}

Formally, we define the \deft{satisfiability problem} of a fact $F$ and constraints
$\Theta$ as checking whether there is an interpretation of 
$\Theta$ and of the existential closure of~$F$.
We will show that the satisfiability problem is undecidable,
not for $\fone \wedge \gctwo$, but for the
weaker $\fone \wedge \alcf$. The DL $\alcf$ is $\gctwo$-expressible; in addition
to the constructors of Section~\ref{sec:prelim}, it also allows
\deft{disjunction} of concepts: $C_1 \sqcup \cdots \sqcup C_n$.

We use tiling systems, following the notations of~\cite{pratt2009data}.
Let $T = (\mathbf{C}, H, V)$ be a tiling system where $\mathbf{C} = C_1, \ldots,
C_N$ is a non-empty finite set of tiles and $H, V \subseteq C^2$ are binary
relations (intuitively standing for ``horizontal'' and ``vertical'').

Given a sequence $\mathbf{c} = c_0, c_1, \ldots, c_n$, the \deft{infinite tiling
problem} for $\mathbf{c}$ is to determine whether there exists an \deft{infinite
tiling}, that is, a function $f : \NN^2 \to C$ such that $f(i, 0) = c_i$ for $0
\leq i \leq n$ and for all $i, j \in \NN$, $(f(i, j), f(i+1, j)) \in H$ and
$(f(i, j), f(i, j+1)) \in V$. It is known that we can choose a fixed $T$ such
that the infinite tiling problem that has $\mathbf{c}$ as input is undecidable.
Hence, fix such a $T$ in what follows.

We consider the (single) $\foneldclass$ rule:
\[
  \tau: \forall u ~ (S(u) \rightarrow  \exists x y z ~ R(u, x) \wedge D(u, y)
  \wedge R(y, z) \wedge D(x, z) \wedge S(x) \wedge S(y) \wedge S(z))
\]
We impose the functionality restrictions $\funct(R)$ and $\funct(D)$.
Intuitively, $R$ stands for ``right'' and $D$ for ``down''.

We create one concept $C_i$ for each tile in $C$. We impose the disjointness
assertions $C_i \sqcap C_j \sqsubseteq \false$ for all $i \neq j$.

We impose the concept inclusions $S \sqsubseteq C_1 \sqcup \cdots \sqcup C_N$.

We impose the concept inclusions $C_i \sqsubseteq \exists R ~ C_{j_1} \sqcup \cdots
\sqcup \exists R ~ C_{j_l}$ where $C_{j_1}, \ldots, C_{j_l}$ are all the tiles
such that $H = \{(C_i, C_{j_k}) \mid 1 \leq k \leq l\}$. Having done this for
$R$ and $H$, we do the same with $D$ and $V$.

We are now ready to conclude the reduction. We claim that the infinite tiling
problem for $T$ and the input $\mathbf{c}$ reduces to the satisfiability of
the fact $F_{\mathbf{c}}$ and the constraints that we have imposed,
where we define:

\[
  F_{\mathbf{c}}(x_0, \ldots, x_n) \defeq S(x_0) \wedge \bigwedge_{0 \leq i \leq n} C_{c_i}(x_i)
  \wedge \bigwedge_{0, \leq i < n} R(x_i, x_{i+1})
\]

Let us prove that, indeed, $F_{\mathbf{c}}$ and the constraints are satisfiable
iff the infinite problem for $T$ with input $\mathbf{c}$ has a solution.

Assume that the infinite tiling problem for $T$ and $\mathbf{c}$ has a solution $f$. Consider the
interpretation $\calI$ such that $\dom(\calI) = \{a_{i,j} \mid i, j \in \NN\}$,
defined as follows:

\begin{align*}
  S^{\calI} & \defeq \{(a_{i,j}) \mid i, j \in \NN\}\\
  R^{\calI} & \defeq \{(a_{i,j}, a_{i+1,j}) \mid i,j \in \NN\}\\
  D^{\calI} & \defeq \{(a_{i,j}, a_{i,j+1}) \mid i,j \in \NN\}\\
  C_{k}^{\calI} & \defeq \{(a_{i,j}) \mid i, j \in \NN, f(i,j) = k\}~\text{for~all}~1 \leq k \leq N
\end{align*}

The interpretation $\calI$ satisfies the rule $\tau$, the disjointness
assertion, the concept inclusions (this uses the fact that $f$ is a tiling
for~$T$), and
the existential closure of $F_{\mathbf{c}}$, so the fact and
constraints are satisfiable.

Conversely, let $\calI$ be an interpretation satisfying the constraints and the
existential closure of $F_{\mathbf{c}}$.
From the fact that $\calI$ satisfies the existential closure of $F_{\mathbf{c}}$,
as $\calI$ satisfies $\tau$ and the two functionality assertions, we can build
from $\calI$ an infinite grid of $R$ and $D$ edges whose top left corner is a
match of variable $x_0$ of~$F_{\mathbf{c}}$,
such that all vertices are in~$S^{\calI}$. Let us index the elements of this grid as
$a_{i,j}$. The constraints impose that each $a_{i,j}$ carries exactly one tile,
so we can define a function $f : \NN^2 \to C$ that maps $(i, j)$ to the one
$C_i$ such that $a_{i,j} \in C_i^{\calI}$ holds. The constraints ensure that $f$ is a
valid tiling for $T$, so the infinite tiling problem for $T$ and $\mathbf{c}$
has a solution.

This concludes the proof that the reduction is correct, so from the
undecidability of the tiling problem we deduce the undecidability of
satisfiability for a fact, a $\foneldclass$ rule, and constraints in $\alcf$. This
implies the claim of Theorem~\ref{thm:fronedl}.

\section{Proofs for Section~\ref{sec:decide}: From Existential Rules to Arity-Two}

\subsection{Proof of Lemma~\ref{lem:fonetogctwo}: Shreddings of $\fonehbasafe$
are in $\gctwo$}

\begin{definition}
  \label{def:cyclef2}
  Recall Definition~\ref{def:cyclef}: we call a $\sigmas$-interpretation $\calJ$
  \deft{cycle-free} if the Gaifman graph $\gaifman{\calJ}$ of $\calJ$ is acyclic.
  
  We call a frontier-one existential rule on $\sigmas$ \deft{cycle-free} if
  the conjunctions of atoms of its head and body are cycle-free.

  We call a CQ $q$ cycle-free if its Gaifman graph is acyclic, defining the
  Gaifman graph $\gaifman{q}$ to have the variables of~$q$ as vertices and an
  edge between any pair of variables that co-occur in an atom of~$q$.
\end{definition}

We first show the following:

\begin{lemma}
  \label{lem:fronegctwook}
  Cycle-free frontier-one existential rules on $\sigmas$ can be translated in
  PTIME to an equivalent $\gctwo$ sentence.
\end{lemma}

The above claim is clearly implied by the following:

\begin{lemma}
  \label{lem:cqgctwook}
  For any cycle-free CQ $q(x)$ on $\sigmas$ with one free variable, $q(x)$ can be translated in quadratic time to an equivalent
  $\gctwo$ formula with one free variable.
\end{lemma}

Indeed, once Lemma~\ref{lem:cqgctwook} is proven, we can show
Lemma~\ref{lem:fronegctwook} by writing the existential rule
\[
  \forall x_\f \mathbf{x} ~ (\phi(x_\f, \mathbf{x}) \rightarrow \exists \mathbf{y}
  \phi(x_\f, \mathbf{y}))
\]
as the following, in $\gctwo$:
\[
  \forall x_\f ~ (\phi'(x_\f) \rightarrow \psi'(x_\f))
\]
where $\phi'$ and $\psi'$ are the formulas obtained from $\phi$ and $\psi$.

Let us then show Lemma~\ref{lem:cqgctwook}:

\begin{proof}
  We test in PTIME whether
  $\gaifman{q}$ is connected. If it isn't,
  we can rewrite $q(x)$ in PTIME as $q'(x) \wedge \bigwedge_i \exists y~ q_i(y)$ where $q'$
  and $q_i$ are CQs whose conjunction of atoms is connected, and translate
  $q(x)$ in PTIME by translating each of the $q_i$. Hence,
  we assume without loss of generality that $\gaifman{q}$ is connected.
  
  We proceed
  by induction on $\card{q}$, the number of atoms of~$q$. If $\card{q} = 1$ the result is trivial. Otherwise, let
  $\calA$ be the set of atoms of $q$ in which the free variable $x$ occurs. Let $\calX$ be the set of
  variables occurring in $\calA$ except $x$. For any $y \in \calX$, let $\calX_y$ be the set
  of variables $z$ different from $x$ and $y$ such that there exists a path from
  $z$ to $y$ in $\gaifman{q}$ which does not go through the vertex $x$. Let
  $\calA_y$ be the set of the atoms of $q$ which are not in $\calA$ and contain a variable of
  $\{y\} \cup \calX_y$. All of these sets can be computed in linear time as the answers to
  reachability questions on $\gaifman{q}$, and the number of sets is linear, so
  the computation takes at most quadratic time.

  We now claim that $\{x\}$, $\calX$, and the $\calX_y$ for $y \in \calX$ are a partition
  of the variables of $x$. Indeed, as $\gaifman{q}$ is connected, any variable
  $z$
  different from $x$ is either adjacent to~$x$ (and thus $z \in \calX$), or there
  is a path from $x$ to it, and the first variable of that path after~$x$ must
  be some $y \in \calX$ (so that $z \in \calX_z$); this justifies that these sets
  cover the variables of $y$. Further, these sets are pairwise disjoint. Indeed, first, $x \notin \calX$ and $x
  \notin \calX_y$ for all $y \in \calX$ by construction. Second, if there is a
  variable $z \in \calX \cap \calX_y$ for some $y \in \calX$, we have $y \neq z$
  as $z \in \calX_y$, and considering the edges in $\gaifman{q}$ between $x$ and
  $y$, $x$ and $z$, and the path from $z$ to $y$ that does not go through $x$,
  we have a cycle in $\gaifman{q}$, a contradiction. Third, for $y, y' \in \calX$, $y
  \neq y'$, if $\calX_y$ and $\calX_{y'}$ are not disjoint, letting $z
  \in \calX_y \cap \calX_{y'}$, as $x$ and $y$, $x$ and $y'$ are connected in
  $\gaifman{q}$, and there is a path from $z$ to $y$ and $z$ to $y'$ in
  $\gaifman{q}$ not going through $x$, we have a cycle in $\gaifman{q}$, a
  contradiction. For similar reasons, $\calA$ and the $\calA_y$ are a partition of the atoms of $q$.

  Now observe that, for any $y \in \calX$, $\calA_y$ is a conjunction of atoms
  with free variables $y$ and $\calX_y$, and
  $\gaifman{\calA_y}$ is acyclic and connected because $\gaifman{q}$ is. Because we have shown
  disjointness, we can apply the induction hypothesis to justify that $\exists
  \calX_y~\calA_y(\calX_y, y)$ can be written in $\gctwo$ as $F_y(y)$, in
  quadratic time in $\exists \calX_y~\calA_y(\calX_y, y)$. Hence, partitioning
  $\calA$ as $\calA'_x$ (the atoms where only $x$ occurs) and 
  $\calA'_y$ for $y \in \calX$ (the atoms of $\calA$ where
  variable $y$ occurs, and the other variable is necessarily $x$), we can
  express $q(x)$ 
  as follows in $\gctwo$:
  \[
  \bigwedge_{A \in \calA_x'} A(x) \wedge \bigwedge_{y \in \calX} \left(\exists
    z~\left(F_y(z) \wedge \bigwedge_{A \in \calA_y'} A(x, z)\right)\right)
\]
  Hence, the overall complexity of the rewriting is quadratic, as the induction
  hypothesis is applied to sets of atoms that are a partition of the atoms of the
  original input formula, so that the quadratic time spent rewriting each set of
  atoms is quadratic overall in the input formula.
  By induction, the proof is completed.
\end{proof}

We then conclude the proof of Lemma~\ref{lem:fonetogctwo} by observing that for
any $\fonehba$ rule~$\tau$, $\shred(\tau)$ is indeed a cycle-free frontier-one
existential rule on $\sigmas$. Indeed, we show this for the head and body with
the following lemma:

\begin{lemma}
  \label{lem:nlcf}
  For any non-looping conjunction of atoms $\Phi$, $\shred(\Phi)$ is cycle-free.
\end{lemma}

\begin{proofb}
  Any cycle in $\gaifman{\shred(\Phi)}$ clearly translates to a Berge cycle in
  $\Phi$ that has length $>2$ or contains a higher-arity atom. In either case,
this would contradict the fact that $\Phi$ is non-looping.
\end{proofb}

\subsection{Proof of Proposition~\ref{prp:fonehba}: QA through shredding}
We start by defining \deft{shreddings} of interpretations:

\begin{definition}
  For any $\sigma$-interpretation $\calI$, the \deft{shredding} $\shred(\calI)$ of
  $\calI$ is the $\sigmas$-interpretation $\calJ$ such that $R^{\calJ} \defeq
  R^{\calI}$ for all $R \in \sigma_{\leq 2}$, $\elt^{\calJ} = \dom(\calI)$, and
  for every $R \in \sigma_{>2}$, for each tuple $\mathbf{a} \in R^{\calI}$, we
  create a fresh element $t \in \dom(\calJ)$, we add $t$ to~$A_R^{\calJ}$, and
  we add $(t, a_i)$ to $R_i^{\calJ}$ for all $1 \leq i \leq \arity{R}$.
\end{definition}

It is immediate that for any $\sigma$-interpretation $\calI$, its shredding
$\calJ$ satisfies $\wf(\sigma)$, and that the unshredding of $\calJ$ (in the
sense of Definition~\ref{def:unshredint}) is $\calI$.

We first show the following lemma to show that negations of CQs, facts and
existential rules are preserved by shredding.

\begin{lemma}
  \label{lem:muchpreserved}
  For every fact~$F$, CQ $q$ and set $\Delta$ of existential
  rules, for any interpretation $\calI$, $\calI$ satisfies $\Delta$, $\neg q$
  and the existential closure of $F$ iff $\shred(\calI)$ satisfies
  $\shred(\Delta)$, $\neg \shred(q)$, and the
  existential closure of $\shred(F)$.
\end{lemma}

To show this, we define the notion of \deft{homomorphism}:

\begin{definition}
  For any interpretations $\calI$ and $\calI'$, a mapping $h : \dom(\calI) \to
  \dom(\calI')$ is a \deft{homomorphism} from $\calI$ to $\calI'$ if for every
  relation $R \in \sigma$, for any tuple $\mathbf{a} \in R^{\calI}$, the tuple
  $h(\mathbf{a}) = (h(a_1), \ldots, h(a_{\arity{R}}))$ is in $R^{\calI'}$.

  This notion extends to homomorphisms from queries to interpretations in the
  usual manner.
\end{definition}

\begin{lemma}
  \label{lem:homreify}
  For any two interpretations $\calI$ and $\calI'$, any homomorphism from
  $\calI$ to $\calI'$ 
  can be extended to a homomorphism from
  $\shred(\calI)$ to $\shred(\calI')$, and conversely any homomorphism from
  $\shred(\calI)$
  to $\shred(\calI')$ can be restricted to a homomorphism from $\calI$ to $\calI'$.
\end{lemma}

\begin{proofb}
  This is immediate, paying attention to the fact that a homomorphism $h$ from
  $\calI$ to $\calI'$ defines a mapping from the tuples of~$\calI$ to the tuples
  of~$\calI'$,
  which describes how to extend $h$ to a homomorphism from $\shred(\calI)$ to
  $\shred(\calI')$ by defining the image of~$h$ on $\dom(\shred(\calI)) \backslash
  \dom(\calI)$.

  Conversely, given a homomorphism from $\shred(\calI)$ to $\shred(\calI')$, its
  restriction to $\dom(\calI)$ is easily seen to be a homomorphism from~$\calI$ to~$\calI'$.
\end{proofb}

We now prove Lemma~\ref{lem:muchpreserved}.

\begin{proof}
  We prove each part of the claim:
\begin{description}
  \item[Query $q$.] By Lemma~\ref{lem:homreify}, there is a homomorphism from
    $\shred(q)$ to $\shred(\calI)$ iff there is a homomorphism from $q$ to
    $\calI$.
  \item[Fact $F$.] Similar to the case of the query.
  \item[Rules $\Delta$.] Consider any existential rule $\tau \in \Delta$.
    
    Assume that $\calI \models \tau$. Consider a homomorphism
    $h$ from the body of~$\shred(\tau)$ (which is the shredding of the body of
    $\tau$) to $\shred(\calI)$, and show that the image of $h$ is not a
    violation of $\tau$ in $\shred(\calI)$. By
    Lemma~\ref{lem:homreify}, $h$ can be restricted to a homomorphism $h'$ from
    the body of $\tau$ to $\calI$. Hence, because $\calI \models \tau$, $h'$ can be
    extended to a homomorphism $h''$ from the body and head of $\tau$ to
    $\calI$. By Lemma~\ref{lem:homreify}, $h''$ can be extended to a homomorphism
    $h'''$ from the shredding of the head and body of $\tau$ to $\shred(\calI)$ that
    matches $h$ on the body of $\shred(\tau)$. So
    we conclude that $h$ does not witness a violation of $\shred(\tau)$.

    Conversely, assume that $\shred(\calI) \models \shred(\tau)$. Consider a
    homomorphism $h$ from the body of $\tau$ to $\calI$. As previously $h$ can be
    extended to a homomorphism $h'$ from the body of $\shred(\tau)$ 
    to $\shred(\calI)$, which can be extended to a
    homomorphism $h''$ from the body and head of $\shred(\tau)$ to
    $\shred(\calI)$. Again we use Lemma~\ref{lem:homreify} to justify that this
    defines a homomorphism $h'''$ from the body and head of $\tau$ to $\calI$ that
    matches $h$ on the body of $\tau$, and conclude that $h$ does not witness a
    violation of $\tau$.\qedhere
\end{description}
\end{proof}

Having proved Lemma~\ref{lem:muchpreserved}, we show the preservation of
$\gctwo$ constraints:

\begin{lemma}
  \label{lem:gctwopreserved}
  For every interpretation $\calI$ and $\gctwo$ theory $\Sigma$, we have $\calI
  \models \Sigma$ iff $\shred(\calI) \models \Sigma$.
\end{lemma}

\begin{proofb}
  The restrictions $\restr{\calI}{\sigma_{\leq 2}}$ and $\restr{\shred(\calI)}{\sigma_{\leq 2}}$ 
  of $\calI$ and $\shred(\calI)$ to $\sigma_{\leq 2}$ are identical (remember that the
  $R_i$ in $\sigmas \backslash \sigma$ are fresh so they do not occur in
  $\Sigma$), hence $\calI$ and $\shred(\calI)$ satisfy the same $\gctwo$ constraints.
\end{proofb}

We can now prove one direction of the result: if there is a
counterexample interpretation of $(\exists \mathbf{x} ~ F(\mathbf{x})) \wedge \Sigma \wedge \Delta \wedge \neg q$, its
shredding is an interpretation of $(\exists \mathbf{x} \mathbf{t} ~
\shred(F)(\mathbf{x}, \mathbf{t})) \wedge \shred(\Delta) \wedge \neg \shred(q)$
(Lemma~\ref{lem:muchpreserved}) that satisfies $\Sigma$
(Lemma~\ref{lem:gctwopreserved}) and $\wf(\sigmas)$ (by our initial immediate
observation about the shredding of interpretations).

What remains is to prove the converse direction of decoding an interpretation
$\calJ$ of
$\Theta \defeq (\exists \mathbf{x} \mathbf{t} ~ \shred(F)(\mathbf{x},
\mathbf{t})) \wedge \Sigma \wedge \shred(\Delta) \wedge \wf(\sigmas) \wedge \neg
\shred(q)$. This is harder, because we must argue that $\calJ$ can be understood as
the shredding of a $\sigma$-interpretation for the above results to apply. This requires us to deal with the issue of
\deft{redundant tuples}:

\begin{definition}
  A $\sigmas$-interpretation $\calJ$ is \deft{redundancy-free} if there is
  no $R \in \sigma_{>2}$, no $t \neq t'$ in $\dom(\calJ)$, and no $\arity{R}$-tuple $\mathbf{a}$ such that
  $(t, a_i)$ and $(t', a_i)$ belong to $R_i^{\calJ}$ for all $1 \leq i \leq \arity{R}$.
\end{definition}

Redundant tuples are the only obstacle to that prevents us from understanding any
interpretation of $\wf(\sigmas)$ as the shredding of some $\sigma$-interpretation. Indeed:

\begin{lemma}
  \label{lem:reifybij}
  $\shred$ is a bijection from $\sigma$-interpretations to redundancy-free
  $\sigmas$-interpretations satisfying $\wf(\sigmas)$.
\end{lemma}

\begin{proofb}
  This is clear, as, writing $\shred^{-1}$ the unshredding operation of
  Definition~\ref{def:unshredint}, we have already observed that, for any
  $\sigma$-interpretation $\calI$, we have $(\shred^{-1} \circ \shred)(\calI) =
  \calI$. Further, given a redundancy-free $\sigmas$-interpretation $\calJ$ satisfying
  $\wf(\sigmas)$, it is immediate that $(\shred \circ \shred^{-1})(\calJ) =
  \calJ$. This concludes the proof.
\end{proofb}

As redundancy-freeness cannot be expressed in $\gctwo$,
our counterexample interpretation
$\calJ$ may not satisfy it. But this does not matter. Recalling our definition of
$\Theta$ above, we show:

\begin{lemma}
  \label{lem:redfree}
  If $\Theta$ has an interpretation then it has a redundancy-free interpretation.
\end{lemma}

\begin{proofb}
  Let $\calJ$ be a $\sigmas$-interpretation of $\Theta$.

  Define the equivalence relation $\sim^{\calJ}$ on $\dom(\calJ)$ as follows: $t
  \sim^{\calJ} t'$
  if, for some $R \in \sigma_{>2}$, $(t)$ and $(t')$ are both in $A_R^{\calJ}$, and for
  every $1 \leq i \leq \arity{R}$, $(t, z) \in R_i^{\calJ}$ iff $(t', z) \in
  R_i^{\calJ}$. The conditions of $\wf(\sigmas)$ ensure that this is an
  equivalence relation, because the $A_R^{\calJ}$ are pairwise disjoint. Define
  $\chi^{\calJ} :
  \dom({\calJ}) \rightarrow \quot{\dom(\calJ)}{\sim^{\calJ}}$ the function  mapping every element of
  $\dom({\calJ})$ to its $\sim^{\calJ}$-equivalence class, and let ${\calJ}'$ be
  the image of ${\calJ}$ under
  $\chi \defeq \chi^{\calJ}$.

  ${\calJ}'$ is redundancy-free as any $t, t' \in \dom({\calJ}')$ witnessing redundancy in
  ${\calJ}'$ would have as preimage by $\chi$ two elements of ${\calJ}$ that are
  $\sim^{\calJ}$-equivalent. (This uses the fact that, by $\wf(\sigmas)$, two
  elements of $A_R^{\calJ}$ and $A_{R'}^{\calJ}$ cannot be adjacent in
  $\gaifman{\calJ}$ for any $R, R' \in  \sigma_{> 2}$.)

  It is easily checked that ${\calJ}'$ is still an interpretation of $\wf(\sigmas)$. As
  $\chi$ is a homomorphism from $\calJ$ to $\calJ'$, and $\calJ$ satisfies the
  existential closure of $F$, $\calJ'$ also satisfies it. Further, because the
  restrictions of ${\calJ}'$ and ${\calJ}$ to $\sigma_{\leq 2}$ coincide,
  ${\calJ}'$ is still an
  interpretation of $\Sigma$.

  To show that ${\calJ}'$ still satisfies $\neg \shred(q)$, it
  suffices to show the existence of a homomorphism from ${\calJ}'$ to
  ${\calJ}$. We build such a homomorphism $h$ by setting, for all $a \in
  \dom({\calJ}')$,
  $h(a) \defeq a'$ for any preimage $a'$ of $a$ by $\chi$. To see why $h$ is a homomorphism,
  consider any tuple $t \in R^{\calJ'}$ for some $R \in \sigmas$. Let $t' \in
  R^{\calJ}$ be a
  preimage of the tuple $t$ by $\chi$.
  Clearly, by $\wf(\sigmas)$, unless $R$ is one of the fresh binary relations
  $R_i$, all elements of $t'$ are singletons in
  their $\sim^{\calJ}$-class, so that necessarily $t' = h(t)$ and $h(t) \in R^{\calJ}$. If $R =
  R_i$, write $t = (u, a)$ and $t' = (u', a')$. We then have that $h(t)$
  is a pair $(u'', a'')$, and necessarily $a'' = a'$ because by $\wf(\sigmas)$ we
  know that $(a') \in \elt^{\calJ}$ so $a' \sim^{\calJ} a''$ implies $a' = a''$.
  Now, as $u' \sim^{\calJ} u''$ as $u', u'' \in A_R^{\calJ}$, and the $A_R$ are
  pairwise disjoint by~$\wf(\sigmas)$, we have $(u', a') \in R_i^{\calJ}$ iff
  $(u'', a'') \in R_i^{\calJ}$, so that indeed $h(t) \in R_i^{\calJ}$.
  Hence, $h$ is indeed a homomorphism from ${\calJ}'$ to
  ${\calJ}$. Thus ${\calJ}'$ satisfies
  $\neg \shred(q)$ because ${\calJ}$ is.

  For any existential rule $\tau$, to show that ${\calJ}'$ still satisfies $\shred(\tau)$, it suffices to observe
  that $h \circ \chi$ is the identity, so that any match $m$ of the body of $\tau$ in
  $\shred(\tau)$ gives such a match in ${\calJ}$ which, as ${\calJ} \models \shred(\tau)$, extends
  to a match of the body and head which is mapped back by $f$ to a match of the
  body and head of $\shred(\tau)$ in ${\calJ}'$, so that $m$ does not witness a
  violation of $\shred(\tau)$. Hence, ${\calJ}'$ still satisfies $\shred(\Delta)$.
\end{proofb}

We can now complete the proof of Proposition~\ref{prp:fonehba} with the
backwards direction: given our interpretation ${\calJ}$ of $\Theta$, make it redundancy-free by
Lemma~\ref{lem:redfree}, and now unshred it
to an interpretation $\calI'$ such that, by Lemma~\ref{lem:reifybij}, $\shred(\calI') = \calJ$. We conclude by
Lemma~\ref{lem:muchpreserved} and Lemma~\ref{lem:gctwopreserved} that $\calI'$
satisfies $\Delta$, $\Sigma$, $\neg q$, and the existential closure
of~$F$.

\subsection{Proof of Lemma~\ref{lem:unravel}: Unraveling for $\gctwo$}

We present the formal unraveling process. In all of this section, we work only
on the signature $\sigmas$.

\begin{definition}
  For any interpretation $\calJ$,
  the \deft{induced interpretation} $\restr{\calJ}{\mathbf{a}}$ of $\calJ$ by $\mathbf{a}
  \subseteq \dom(\calJ)$ is the interpretation containing all the tuples of $\calJ$
  where only elements of $\mathbf{a}$ occur. A \deft{guarded pair} in $\calJ$ is a
  pair $\{a, b\}$ of two distinct elements of $\dom(\calJ)$ such that $a$ and $b$
  co-occur in some tuple of $\calJ$. The \deft{immediate neighborhood}
  $\ineig{\calJ}{a}$
  of $a \in \dom(\calJ)$ in $\calJ$ is $\{b \in \dom(\calJ) \backslash \{a\} \mid \{a, b\}
  \text{~guarded pair in~}\calJ\}$.

  The \deft{bags} of an interpretation $\calJ$
  are the interpretations induced by all guarded pairs of $\calJ$.
  The \deft{bag graph} of a $\sigmas$-interpretation $\calJ$ is the
  undirected graph on the bags of $\calJ$ (without self-loops) where two
  distinct bags are
  adjacent whenever their domains share one common element. (As the domains have
  size two, they must then share exactly one element.)

  Given a witness $\calW$ of a fact $F$ in $\calJ$, 
  we alter the definition of the bag graph of $\calJ$ by adding one \deft{fact bag} corresponding to
  the witness $\calW$; the fact bag is adjacent to
  all bags with which it shares one element (but not those with which it shares
  two elements).
\end{definition}

\begin{definition}
  \label{def:unraveling}
  A \deft{tree-like interpretation} is a tree $T = (W, E,
  b_\r)$ where each $b \in W$ is a bag (that is, an interpretation), $b_\r \in W$ is
  the root bag, and $E$ is the directed edge relation. We require that for all $(b, b')
  \in E$, the domains of $b$ and $b'$ share exactly one element $u$ such that
  $u$ exactly occurs in $T$ at the following places: in $b$ (we say it was
  \deft{introduced} in $b$), and in all children of $b$ (including $b'$).
  Further, if two bags $b$ and $b'$ in $W$ share
  some element then either they are siblings in $T$ or one is a child of the other in
  $T$.
  We write $\dom(T) = \bigcup_{b \in W} \dom(b)$ and
  also see $T$ as the interpretation $\bigcup_{b \in W} b$.

  Given a fact $F$ and a witness $\calW$ of $F$ in $\calJ$, we say that $T$ is
  an \deft{unraveling} of $\calJ$ preserving $\calW$ if 
  $b_\r$ is the fact bag of $\calJ$, all other bags of $T$ have domain of
  size~$2$, and elements of $\dom(\calJ_0)$ only occur in $b_\r$ (we say they were
  \deft{introduced} in $b_\r$).
\end{definition}

Our goal will be to construct an unraveling of the counterexample interpretation, because
of the following:

\begin{lemma}
  \label{lem:cfree}
  If $T$ is an unraveling of an interpretation $\calJ$ preserving a witness
  $\calW$ of a fact $F$, then $T$ is an
  interpretation where $\calW$ is also a witness of~$F$ and which is cycle-free
  except for~$\calW$ (recall Definition~\ref{def:cyclef}).
\end{lemma}

\begin{proofb}
  Except for $\dom(\calW)$, $\gaifman{T}$ is a tree which matches $T$: if any two
  elements $u, v$ of $\dom(T)$ are not both in $\dom(\calW)$ and co-occur in a
  tuple of
  $T$, this edge of $\gaifman{T}$ corresponds to the edge between the bag where
  $u$ was introduced, and the bag where $v$ was introduced.
\end{proofb}

However, we also want the unraveling to be \deft{faithful}, so the constraints
are preserved.

\begin{definition}
  $T = (W, E, b_\r)$ is a \deft{faithful} unraveling of an interpretation
  $\calJ$ preserving $\calW$ (where $\calW$ is a witness of a fact $F$) if it is
  an unraveling of $\calJ$ preserving $\calW$ such that there exists a homomorphism $\pi$ from
  $T$ to
  $\calJ$, and a mapping $\phi$ from $W$ to the bags of $\calJ$ that maps $b_\r$
  to the fact bag, and maps no other bag to the fact bag. We require
  that:
  \begin{description}
    \item[(Compat)] $\phi$ is \deft{compatible} with $\pi$: for any $b \in W$, 
      $\restr{\pi}{\dom(b)}$ is an isomorphism between $b$ and $\phi(b)$, and it is
      even the identity for $b = b_\r$;
    \item[(IN)] for every $a \in \dom(T)$, $\restr{\pi}{\ineig{T}{a}}$ is an
      isomorphism between $\ineig{T}{a}$ and $\ineig{\calJ}{\pi(a)}$;
    \item[(Surj)] $\phi$ is \deft{surjective} except for $\calW$: for any bag
      $b$ of $\calJ$ whose
      domain is not a subset of $\dom(\calW)$, $b$ has a preimage by $\phi$.
  \end{description}
\end{definition}

We say an interpretation is \deft{unravelable} if all elements of the
interpretation occur
in at least one tuple for a binary relation, and if its bag graph is connected; we can assume
without loss of generality that interpretations are unravelable by adding tuples for a
fresh binary relation to satisfy these conditions. We now claim:

\begin{proposition}
  \label{prp:constraintsok}
  For any fact $F$, $\gctwo$ constraints $\Theta$, and CQ $q'$, 
  if $\calJ$ is an unravelable interpretation that satisfies $\Theta$ and
  $\neg q'$ and has a witness $\calW$ of~$F$, and
  $T$ is a faithful unraveling of $\calJ$ preserving $\calW$, then $T$ (seen as an interpretation)
  satisfies
  $\Theta$, $\neg q'$, and the existential closure of~$F$ (in fact $\calW$ is
  still a witness of~$F$ in $T$).
\end{proposition}

\begin{proofb}
  It is clear that $\calW$ is still a witness of $F$ in $T$.
  $T$ also satisfies $\neg q'$, since  there exists a homomorphism
  from $T$ to $\calJ$, so if $T$ satisfied $q'$ then so would $\calJ$.

  We must show that $T$ still satisfies $\Theta$. Up to  expanding the
  original interpretation by interpreting new relation names, following
  \cite{kazakov2004polynomial} we can rewrite the $\gctwo$ constraints $\Theta$
  as a conjunction of a formula of $\gftwo$ (the guarded fragment with two
  variables but no number restrictions) and number restrictions of the form
  $\forall x ~ \exists^{\bowtie n} y ~ R(x, y)$ where $n \in \mathbb{N}$,
  $\bowtie\,\, \in \{\geq, <\}$, and $R$ is a binary relation.

  The fact that the number restrictions are preserved is immediate, since they
  only depend on the immediate neighborhood of elements, which are isomorphically
  preserved by $\pi$ according to property (IN).

  We show that $\gftwo$ is preserved by showing the existence of a guarded
  bisimulation from $T$ to $\calJ$ \cite{gradel2002back}. We define the guarded
  bisimulation as the set $\calI$ of all restrictions of $\pi$ to singletons and guarded pairs of
  $T$, which are indeed partial isomorphisms from $T$ to $\calJ$. We show that
  the back and forth conditions are satisfied. For any $f : X \to Y$ in $\calI$:

  \begin{description}
    \item[Forth.] Consider a guarded set $Z$ of $T$. There is a partial
      isomorphism $f'$ in $\calI$ with domain $Z$, and it
      agrees with $f$ on $Z \cap X$ as they are both restrictions of $\pi$
    \item[Back.] Consider a guarded set $Z$ of $\calJ$. As $\calJ$ is unravelable, all 
      singletons of $\calJ$ occur in some guarded pair of $\calJ$, so it
      suffices to consider the case where $\card{Z} = 2$.
      Let $b$ be the corresponding bag of $\calJ$. We distinguish depending on
      whether $Z$ does not intersect $Y$ or whether it does:
  
      If $\card{Z \cap Y} = 0$, either $\dom(b) \subseteq \dom(\calW)$ so we can find
      an isomorphism of $\calI$ with domain
      $\pi^{-1}(Z)$ because $\pi$ is the identity on $\dom(\calW)$,
      or as $\phi$ is surjective (property (Surj)) there exists $b' \in
      W$ such that $\phi(b') = b$ and thus, because $\phi$ and $\pi$ are
      compatible by property (Compat),
      the image of $\restr{\pi}{\dom(b')}$ is $Z$, so there is a corresponding
      partial isomorphism in $\calI$.
  
      If $\card{Z \cap Y} \neq 0$, the only non-trivial case is $\card{Z \cap Y} = 1$.
      Let $a$ be the element of $Z \cap Y$. Because by property (IN)
      $\restr{\pi}{\ineig{T}{a}}$ is an isomorphism from
      $\ineig{T}{a}$ to $\ineig{\calJ}{\pi(a)}$, there exists a guarded pair
      $X'$ of $T$ such that $\pi(X') = Z$; hence, there is a partial
      isomorphism $f'$ in $\calJ$ from $X'$ to $Z$, and it agrees with $f$ as both
      are restrictions of $\pi$.
  \end{description}

  This concludes the proof.
\end{proofb}

We must now show that a faithful unraveling exists:

\begin{proposition}
  \label{prp:haveunravel}
  For any fact $F$, for any unravelable interpretation $\calJ$ and witness
  $\calW$ of $F$ in $\calJ$, there is a faithful unraveling $T$ of
  $\calJ$ preserving $\calW$.
\end{proposition}

\begin{proofb}
To build $T$, define the root $b_\r$ of~$T$ as $\calW$, set $\phi(t_\r) =
\calW$,
initialize $\pi$ as the identity on $\calW$, and define inductively $T = (W, E,
t_\r)$, the homomorphism $\pi$ and the mapping $\phi$, as follows. At every bag
$b \in W$,
consider the corresponding bag $\phi(b)$ of~$\calJ$. For every element $a$
introduced in $b$ (there is only one except for $b = b_\r$), consider every bag
$b''$ in the bag graph of~$\calJ$ that shares element $\pi(a)$ with
$\phi(b)$ (so $b''$ is adjacent to $\phi(b)$ in the bag graph). Letting $\dom(b'') = \{\pi(a),
a''\}$, create a bag $b'''$ in $T$ as a child of~$b$, with domain $\{a, a'\}$
where $a'$ is fresh and where we set $\pi(a') \defeq a''$, and make $b'''$ an
isomorphic copy of~$b''$ following the mapping $\pi$. Perform the same process inductively on
all child bags.

It is clear that the result of this process is indeed an unraveling of~$\calJ$. It
is also clear that $\pi$ thus defined is a homomorphism as any created tuple in $T$
clearly has a homomorphic image via $\pi$. Last, it is clear that $\phi$ maps $t_\r$,
and only $t_\r$, to the fact bag. For property (Compat), it is clear that the
restriction of~$\pi$
to any bag~$b$ of $T$ is an isomorphism between $b$ and $\phi(b)$.
For property (IN), for any element $a \in \dom(T)$ it is
clear that $\restr{\pi}{\ineig{T}{a}}$ is an isomorphism from $\ineig{T}{a}$ to
$\ineig{\calJ}{\pi(a)}$: $\ineig{T}{a}$ consists of the union of the bag $b_a$
where $a$ was introduced and the children of $b_a$ with which $a$ is shared (i.e., all
children, except at $b_\r$), which corresponds exactly to the bags of $\calJ$
where $\pi(a)$ occurs.
For property (Surj), the surjectivity of $\phi$ is because $\calJ$ is
unravelable, so all bags of $\calJ$ are reachable from the fact bag.
\end{proofb}

This concludes the proof: we make the interpretation unravelable without loss of
generality, unravel it with Proposition~\ref{prp:haveunravel}, and
Proposition~\ref{prp:constraintsok} and Lemma~\ref{lem:cfree} ensure that the result satisfies the
required conditions.

\subsection{Proof of Lemma~\ref{lem:preserv2}: Treeification soundness}

We call a \deft{bad cycle} in a conjunction of $\sigma$-atoms $\Phi$ a
Berge cycle of length $>2$ or containing a higher-arity atom (following
Definition~\ref{def:berge}).

Let $F$ be a $\sigmas$-fact, $\tau$ be a $\foneha$ rule, and $\calJ$ be a
$\sigmas$-interpretation. Assume that $\calJ$ is cycle-free except for
$\shred(F)$, and let $\calW$ be the witness whose existence is guaranteed by
this.
Similarly to
Lemma~\ref{lem:nlcf}, it is easily seen that this implies that $\calI$ is
non-looping except within the domain of the unshredding $\calW'$ of $\calW$.

Now, assume that $\calJ$ satisfies $\shred(\treeif_F(\tau))$,
and assume that $\calI \not\models
\tau$. Let $f$ be a mapping from the body of $\tau$ to $\calI$ that witnesses the
violation. We consider the dependency $\tau'$ (implied by $\tau$) obtained by
identifying all variables of the body of $\tau$ that are mapped to the same
element by $f$. We can thus see $f$ as a match of $\tau'$ that maps all
variables of the body of $\tau'$ to distinct elements. If $\tau'$ is a
$\fonehba$ rule, then it is in $\treeif_F(\tau)$ (taking $\mathbf{x}' =
\emptyset$), so that if $\calI$ violates $\tau$
then it violates $\treeif_F(\tau)$, contradicting the fact that it is the
unshredding of $\calJ$ which satisfies $\shred(\treeif_F(\tau))$ (as in
Proposition~\ref{prp:fonehba}).

Hence, assume that $\tau'$ is not a
$\fonehba$ rule, so that its body has a bad cycle. Because $f$ maps all
variables in the body of $\tau'$ to distinct elements of $\calI$, the image of any
bad cycle of the body of $\tau'$ by $f$ is a bad cycle of $\calI$. Hence, as
$\calI$ is non-looping except for $\calW'$, any bad cycle of
$\tau'$ must be mapped by $f$ to elements of $\dom(\calW')$. Now consider $\tau''$
obtained from $\tau'$ by setting $\mathbf{x}'$ to be the variables mapped to
elements of $\dom(\calW')$, setting $g$ that maps each variable $x$ of $\mathbf{x'}$
to the variable $z$ of $F$ such that we have $f(x) \in P_z^{\calI}$ (there is precisely one,
as $\calW$ is a witness of $\shred(F)$, which we have defined to include the
atoms $P_\bullet(\bullet)$), and performing the construction $g(\tau'')$ as
in Definition~\ref{def:treeif}. The result $\tau''$ is in $\fonehba$, as otherwise a
bad cycle in it translates to a bad cycle in $\tau'$ of elements not matched to
$\dom(\calW')$, which, as we have seen, contradicts the fact that $\calI$ is non-looping
except within $\dom(\calW')$. So $\tau''$ is in $\treeif_F(\tau)$, and $f$ is also a match of
$\tau''$ that maps the frontier variable to the same element. Hence, as
$\calI \models \treeif_F(\tau)$, we have a contradiction of the fact that~$f$
witnesses a violation.

This concludes the proof.

\section{Proofs for Section~\ref{sec:fds}: Adding Functional Dependencies}

\subsection{Proof of Theorem~\ref{thm:fdsfrone}: QA is undecidable for FDs and
single-head $\fone$ rules}
\label{apx:prf_fdsfrone}

Call $\fonesh$ the class of single-head frontier-one rules. Recall the
definition of the entailment problem (Definition~\ref{def:ent})
and of $\ufdclass$ (Definition~\ref{def:ufd}).  We will write rules of
$\uidclass$ and $\bidclass$ as in Definition~\ref{def:uid}.

By Lemma~\ref{lem:qatoimp}, the entailment problem for $\fonesh \wedge
\ufdclass$ and $\bidclass$ reduces to QA for $\fonesh \wedge \ufdclass$, so it
suffices to show the undecidability of the former to show undecidability of the
latter. We will do so by adapting the result of \cite{mitchell1983implication}, who
showed that implication of $\bidclass$ rules by $\ufdclass$ and
$\bidclass$ constraints
is undecidable.
We will need to consider a special form of the problem studied in \cite{mitchell1983implication}:

\begin{definition}
  The \deft{restricted $\ufdclass/\bidclass$ entailment problem} is the entailment problem for
  $\ufdclass \wedge \bidclass$ and $\bidclass$ where the input is restricted so
  that there is only one relation $R$, and, for any $\bidclass$ rule $R^a
  R^b \subseteq R^c R^d$ in the input, the UFD $R^a \rightarrow R^b$ holds in
  the input.
\end{definition}

We now state our variant of the undecidability result in \cite{mitchell1983implication}:
\begin{theorem}
  \label{thm:mitchell2}
  The restricted $\ufdclass/ \bidclass$ entailment problem is undecidable.
\end{theorem}

\begin{proofb}
  We recall the proof technique of \cite{mitchell1983implication}.
  The proof gives a reduction to the entailment problem
from the following
  undecidable problem: given a system of equations of the form $x = y \circ z$
  on functional monoids, decide if a certain equation $x_0 = y_0 \circ z_0$ is
  entailed by the system.

  This problem is reduced to the entailment problem in the following way. Given such a
  system, we create a relation $R$ with one attribute $R^x$ per variable $x$,
  plus an extra attribute $R^\a$. We impose the UFD $R^\a \rightarrow R^x$
  and the $\uidclass$ rule $R^x \subseteq R^a$ for each position $R^x$ of $R$ (except
  $R^\a$). This ensures that the projection of $R$ to $R^a R^x$ can be
  interpreted as the graph of a function. Now, equations of the form $x = y
  \circ z$ can be understood as the corresponding assertions on the functions
  represented by $R^\a R^x$, $R^\a R^y$ and $R^\a R^z$, and Lemma~4 of
  \cite{mitchell1983implication} shows that such an assertion can actually be
  enforced by a $\bidclass$-like constraint: $R^y R^x \subseteq R^\a R^z$. Those constraints are
  not necessarily $\bidclass$ constraints because we may have $R^x = R^y$.

  We  observe that we can enforce that we always have $x \neq y$ in such
  constraints by adding more
  equations. For every variable $x$, we replace all its occurrence in the
  equations by fresh variables $x_1, \ldots, x_n$, and we add the equations $x_1
  = x_2$, $\ldots$, $x_{n-1} = x_n$. Clearly the resulting problem is equivalent to
  the original one, and the encoding of each constraint $x = y \circ z$ is now
  an actual $\bidclass$ rule. Similarly to Lemma~4 of
  \cite{mitchell1983implication}, we observe that the
  new equations of the form $x_i = x_{i+1}$ are equivalent to asserting $R^a
  R^{x_i} \subseteq R^\a R^{x_{i+1}}$ and $R^\a R^{x_{i+1}} \subseteq R^\a
  R^{x_i}$ on the projections.

  We now observe that the implication problem of
  \cite{mitchell1983implication} with the above restriction can in fact be
  assumed to be in the form of the restricted $\ufdclass$/$\bidclass$ problem,
  except that it features some $\uidclass$ rules.
  Indeed, each of the $\bidclass$ rules in the encoding of the equations $x = y \circ z$ is of the form $\tau: R^y R^x \subseteq
  R^\a R^z$, and the $\ufdclass$ constraint $\phi: R^\a \rightarrow R^z$ holds. It is clear that
  $\tau \wedge \phi \entunr \phi'$, where $\phi': R^y \rightarrow R^x$. Indeed, any
  violation of $\phi'$ in an interpretation satisfying $\tau$ implies by $\tau$ the
  existence of a violation of $\phi$. Hence, the problem is equivalent to the
  one where we add the UFDs $R^y \rightarrow R^x$ for every equation $x = y
  \circ z$. For the equations of the form $x_i = x_{i+1}$, as $R^\a \rightarrow
  R^{x_i}$ and $R^\a \rightarrow R^{x_{i+1}}$ hold, the condition of the
  restricted $\ufdclass$/$\bidclass$ problem is also satisfied.

  The last step to reduce to the restricted $\ufdclass/\bidclass$ setting is to eliminate the
  $\uidclass$ rules. 
  We do this using a variant of Lemma~\ref{lem:rmuid}, where we encode each
  $\uidclass$ rule $\tau: R^p \subseteq
  S^q$ as the $\bidclass$ rule $R^p R^{\tau,1} \subseteq R^q R^{\tau,2}$, where
  $R^{\tau,1}$ and $S^{\tau,2}$ are fresh positions of $R$ and $S$ respectively,
  plus the UFD $R^p \rightarrow R^{\tau,1}$ so that the condition of the
  restricted $\ufdclass/\bidclass$ problem is respected. It is easily seen that this does not affect the
  rest of the proof: projecting away the additional attributes or populating
  them with the same value as their determiner cannot violate any of these
  additional UFDs.
\end{proofb}

What remains now is to show the following:

\begin{proposition}
  \label{prp:foneshtost}
  There is a reduction from the restricted $\ufdclass/\bidclass$ entailment problem to entailment for
  $\ufdclass \wedge \fonesh$ and $\bidclass$.
\end{proposition}

\begin{proofb}
  Consider an instance of the restricted UFD/BID entailment problem: we are
  given a relation~$R$, a set
  $\Phi$ of UFDs, a set $\Delta$ of $\bidclass$ rules, and the $\bidclass$
  rule $\tau$, and we ask whether $\Phi \wedge \Delta \entunr \tau$.

  Let $n$ be the number of positions of $R$.
  We construct the relation $S$ whose positions are $S^{i,1}$ and $S^{i,2}$ for
  every position $R^i$ of $R$. We translate each UFD $\phi:R^p \rightarrow R^q$ of~$\Phi$
  to the two UFDs $\phi_i:S^{p,i} \rightarrow S^{q,i}$ for $i \in \{1, 2\}$,
  letting $\Phi'$ be the resulting UFDs on $S$. We translate the $\bidclass$
  rule $\tau: R^a R^b \subseteq R^c R^d$ to
  the $\bidclass$ rule $\tau': S^{a,1} S^{b,1} \subseteq S^{c,1} S^{d,1}$. We now
  describe how each $\bidclass$ rule of $\Delta$ is translated to $\fonesh$.

  Consider a $\bidclass$ rule $\delta:R^a R^b \subseteq R^c R^d$. We create a first $\fonesh$
  rule
  \[
    \delta_1: \forall \mathbf{x} ~ \left(S(x^1_1, \ldots, x^1_n, x^2_1, \ldots, x^2_n) \rightarrow
    \exists \mathbf{y} ~ S(z^1_1, \ldots, z^1_n, z^2_1, \ldots, z^2_n)\right)
  \]
  defined as follows:
  \begin{itemize}
  \item $z^1_a$ is $x^1_a$;
  \item $z^2_a$ is $x^1_a$;
  \item $z^2_b$ is $y^1_b$;
  \item otherwise, $z^i_j$ is $y^i_j$.
  \end{itemize}

  We create a second $\fonesh$ rule
  \[
    \delta_2: \forall \mathbf{x} ~ \left(S(x^1_1, \ldots, x^1_n, x^2_1, \ldots, x^2_n) \rightarrow
    \exists \mathbf{y} ~ S(z^1_1, \ldots, z^1_n, z^2_1, \ldots, z^2_n)\right)
  \]
  defined as follows:
  \begin{itemize}
  \item $z^2_a$ is $x^2_a$;
  \item $z^1_c$ is $x^2_a$;
  \item $z^1_d$ is $y^2_b$;
  \item otherwise, $z^i_j$ is $y^i_j$.
  \end{itemize}

  For instance, the $\bidclass$ rule $\delta: R^1 R^2 \subseteq R^3 R^4$ would be
  encoded as:
  \begin{align*}
    \delta_1:& \forall \mathbf{x} ~ \left(
    S(x^1_1, x^1_2, x^1_3, x^1_4, x^2_1, x^2_2, x^2_3, x^2_4)
    \rightarrow \exists \mathbf{y} ~
    S(x^1_1, y^1_2, y^1_3, y^1_4, x^1_1, y^1_2, y^2_3, y^2_4)\right)\\
    \delta_2:& \forall \mathbf{x} ~ \left(
    S(x^1_1, x^1_2, x^1_3, x^1_4, x^2_1, x^2_2, x^2_3, x^2_4)
    \rightarrow \exists \mathbf{y} ~
    S(y^1_1, y^1_2, x^2_1, y^2_2, x^2_1, y^2_2, y^2_3, y^2_4)\right)
  \end{align*}
  Note that, by the condition of the restricted UFD/BID entailment problem, the UFD
  $R^1 \rightarrow R^2$ holds in $\Phi$. Hence,
  $y^1_2$ in the head of the first rule must be matched to the same
  element as $x^1_2$, and likewise for $y^2_2$ in the second rule.
  
  We let $\Delta'$ be the result of this encoding of~$\Delta$, and we claim that
  $\Phi \wedge \Delta \entunr \tau$ iff $\Phi' \wedge \Delta' \entunr \tau'$.

  \medskip

  To this end, we first show that, for any $\bidclass$ constraint $\delta: R^a R^b \subseteq R^c
  R^d$ of $\Delta$, with $\phi: R^a \rightarrow R^b$ in $\Phi$
  by the assumption of the restricted UFD/BID
  entailment problem, 
  considering the translations $\phi_1, \phi_2 \in \Phi'$ of~$\phi$, and
  considering and $\delta_1, \delta_2 \in \Delta'$ as
  defined above,
  letting $\delta': S^{a,1} S^{b,1} \subseteq S^{c,1} S^{d,1}$ be the intuitive
  $\bidclass$ translation of $\delta$ to~$S$, the following entailment holds:
  $\delta_1 \wedge \delta_2 \wedge \phi_1 \wedge \phi_2 \entunr \delta'$. In
  other words, our rewriting $\delta_1$ and $\delta_2$ of $\delta$ implies the
  straightforward rewriting $\delta'$.
  
  Indeed, consider an interpretation $\calI$ of $\delta_1 \wedge \delta_2 \wedge
  \phi_1 \wedge \phi_2$. Consider a tuple $t = (u^1_1, \ldots, u^1_n, u^2_1, \ldots,
  u^2_n) \in S^{\calI}$
  We wish to show that it does not witness a violation of $\delta'$.
  By $\delta_1$, there exists a tuple $(v^1_1, \ldots, v^1_n, v^2_1,
  \ldots, v^2_n) \in S^{\calI}$ with $v^1_a = u^1_a$, $v^2_a = u^1_a$, and $v^1_b =
  v^2_b$. As $\calI$ satisfies $\phi_1$, as $v^1_a = u^1_a$, we must have $v^1_b = u^1_b$, so that
  $v^2_b = u^1_b$. Now, by $\delta_2$, there exists a tuple $t' = (w^1_1,
  \ldots, w^1_n, w^2_1, \ldots, w^2_n) \in S^{\calI}$ with $w^2_a = v^2_a$, $w^1_c =
  v^2_a$, and $w^1_d = w^2_b$. Now, as $\calI$ satisfies $\phi_2$, as $w^2_a =
  v^2_a$, we must have $w^2_b = v^2_b$. Putting it together, we have $w^1_c =
  v^2_a = u^1_a$, and $w^1_d = w^2_b = v^2_b = u^1_b$. Hence, $t'$ witnesses
  that $t$ is not a violation of $\delta'$.
  This proves that, indeed, $\delta_1 \wedge \delta_2 \wedge \phi_1 \wedge \phi_2 \entunr \delta'$.

  \medskip

  Let us now proceed with the proof of the fact that
  $\Phi \wedge \Delta \entunr \tau$ iff $\Phi' \wedge \Delta' \entunr \tau'$,
  to show that the reduction is correct. Assume
  that $\Phi' \wedge \Delta' \not\entunr \tau'$. Let $\calI$ be an
  interpretation of $\Phi', \Delta'$ that violates $\tau'$. Let $\calJ$ be the
  projection of $\calI$ to the
  positions $S^{1,1}, \ldots, S^{n,1}$, formally:
  \[
    R^{\calJ} = \{(a_1^1, \ldots, a_n^1) \mid (a_1^1, \ldots, a_n^1, a_1^2, \ldots, a_n^2) \in S^{\calI}\}
  \]
  Because $\calI$ satisfies $\Phi'$, $\calI$
  clearly satisfies $\Phi$. By our previous observation, it is clear that,
  because $\calI$
  satisfies $\Delta'$ and $\Phi$, $\calJ$ satisfies $\Delta$. It is also clear
  that, because $\calI$ violates $\tau'$, $\calJ$ violates $\tau$. So $\calJ$
  witnesses that $\Phi \wedge \Delta \not\entunr \tau$.

  \medskip

  Conversely, assume that $\Phi \wedge \Delta \not\entunr \tau$, and let $\calI$ be a
  counterexample interpretation. We create $\calJ$ by constructing $S$ as the product of $R$
  by itself: create the tuple $(\mathbf{a}, \mathbf{b}) \in S^{\calJ}$ for
  every tuples $\mathbf{a}, \mathbf{b} \in R^{\calI}$. It is clear that $\calJ$ satisfies
  $\Phi'$ because $\calI$ satisfies $\Phi$ (as the FDs are either within the
  positions $S^{i,1}$ or within the positions $S^{i,2}$). For the same reason
  $\calJ$ still violates $\tau'$
  because $\calI$ did. We now check that $\calJ$ satisfies $\Delta'$. Let
  $\delta:R^a R^b \subseteq R^c R^d$
  be
  a rule of $\Delta$ and show that $\calJ$ satisfies $\delta_1$ and $\delta_2$. For
  $\delta_1$, let $t = (\mathbf{u}, \mathbf{v})$ be a tuple of $S^{\calJ}$. By construction
  of $\calJ$ we have $(\mathbf{u}, \mathbf{u}) \in S^{\calJ}$ which witnesses that
  that $F$ is not a violation of $\delta_1$. For $\delta_2$, let $t =
  (\mathbf{u}, \mathbf{v})$ be a tuple of $S^{\calJ}$. By construction of
  $\calJ$ we have $\mathbf{v} \in R^{\calI}$. As $\calI$ satisfies $\delta$,
  there is a tuple $\mathbf{w} \in R^{\calI}$
  such that $w_c = v_a$ and $w_d = v_b$.
  By construction of $\calJ$, we have $(\mathbf{w}, \mathbf{v}) \in S^{\calJ}$, which witnesses that $t$ is not a
  violation of $\delta_2$. Hence $\calJ$ satisfies $\Delta'$, so it witnesses that
  $\Phi' \wedge \Delta' \not\entunr \tau'$.

  This shows that our reduction is sound, and concludes the proof.
\end{proofb}

We conclude the proof of Theorem~\ref{thm:fdsfrone} by combining
Lemma~\ref{lem:qatoimp},
Proposition~\ref{prp:foneshtost}
and Theorem~\ref{thm:mitchell2}.

\subsection{Proof of Lemma~\ref{lem:whysafe2}: FD-safety and cycle-freeness}

Let $\Phi$ be a set of FDs on $\sigma_{>2}$, let $\calJ$ be a
$\sigmas$-interpretation, and assume that it is cycle-free and FD-safe except
for a witness $\calW$ (of some $\sigmas$-fact~$F$). Note that that there is a
slight abuse of terminology here relative to Definition~\ref{def:cyclef}: we mean
that $\calJ$ is cycle-free except for $F$, and that $\calW$ is a witness
satisfying the conditions of the definition of being cycle-free.

Let $\calI$ be the unshredding of $\calJ$, and 
consider two tuples $\mathbf{a}$ and $\mathbf{b}$ in $R^{\calI}$ that violate an
FD $\phi$ of $\Phi$ (remember that this implies $\arity{R} > 2$). By our assumption that the unshredding $\calW'$ of $\calW$
satisfies $\Phi$, it is not possible that both $\mathbf{a}$ and $\mathbf{b}$ are
in $R^{\calW'}$. Let $P$ be the positions of $R$ that are the determiner
of $\phi$, and $R^r$ be the position that $\phi$ determines, so that $a_i = b_i$
for all $R^i \in P$, but $a_r \neq b_r$.

Consider the set $S = \{a_i \mid R^i \in P\}$. If $S$ is not a singleton set,
then, as $\arity{R} > 2$ and $\mathbf{a} \neq \mathbf{b}$, the image of the shredding of $\mathbf{a}$ and $\mathbf{b}$ creates a 
cycle in $\gaifman{\shred(\calI)}$, which does not consist only of elements of $\calW$
because $\mathbf{a}$ and $\mathbf{b}$ are not both in $R^{\calW'}$. This
contradicts the fact that $\calJ$ should be cycle-free except for $\calW$.
Hence, $S$ is a singleton set.

Accordingly, let $a$ be the common element which is the $a_j$ for any $R^j \in
P$.
Now, considering the shredding of $\mathbf{a}$ and
$\mathbf{b}$ in $\calJ' \defeq \shred(\calI)$, $\shred(\calI)$ is such that $(t,
a)$ and $(t', a)$ are in $R_i^{\calI'}$ for all $R^i \in P$.
As $P$ is the determiner of a FD of
$\Phi$, $\shred(\calI)$ is not FD-safe except for $\calW$, because $t$ and
$t'$ cannot both be in $\dom(\calW)$, otherwise $\mathbf{a}$ and $\mathbf{b}$
would be in $R^{\calW'}$. This contradicts the fact that $\calJ$ is FD-safe
except for $\calW$

\subsection{Proof of Lemma~\ref{lem:unravelb}: Unraveling with FDs}

We first assume without loss of generality that the $\foneha$ constraints have
only unary or higher-arity relations in their head. Indeed, for any $\foneha$
rule $\tau$ violating this condition, we can replace its
head by a fresh unary atom $U(x)$, where $x$ is the frontier variable, and
assert in the $\gctwo$ constraints that $U$ implies the head atom of $\tau$.

We first define:

\begin{definition}
  A \deft{proper guarded pair} of a $\sigmas$-interpretation $\calJ$ is a pair $\{a, b\}$
  of distinct elements of $\dom(\calJ)$ such that $a$ and $b$ co-occur in a
  relation which is \emph{not} in $\sigmas \backslash \sigma$.
  Note that if $\calJ$ satisfies $\wf(\sigmas)$ then, for any guarded
  pair, either the pair only occurs in tuples for such relations, or the pair
  only occurs in tuples for relations of $\sigmas \backslash \sigma$.

  The \deft{proper bags} of $\calJ$ are the bags induced by proper guarded pairs.

  Given a $\sigmas$-interpretation $\calJ$ and
  $(a) \in \mathrm{Elt}^{\calJ}$,
  the \deft{arity-two immediate neighborhood} $\ineigtwo{\calJ}{a}$ of $a$ in
  $\calJ$
  is the restriction of $\ineig{\calJ}{a}$ to the proper guarded pairs.
\end{definition}

We give a different name to the unravelings that we will create:

\begin{definition}
  $T = (W, E, b_\r)$ is an \deft{FD-faithful} unraveling of an interpretation
  $\calJ$ preserving a witness $\calW$ given FDs $\Phi$ if it is
  an unraveling of $\calJ$ preserving $\calW$ (recall Definition~\ref{def:unraveling}) such that there exists a homomorphism $\pi$ from
  $\dom(T)$ to
  $\dom(\calJ)$, and a mapping $\phi$ from $W$ to the bags of $\calJ$ that maps $b_\r$
  to the fact bag, and maps no other bag to the fact bag. We require
  that:
  \begin{description}
    \item[(Compat-P)] $\phi$ is \deft{compatible} with $\pi$: for any $b \in W$ such that
      $\phi(b)$ is a proper bag, 
      $\restr{\pi}{\dom(b)}$ is an isomorphism between $b$ and $\phi(b)$, and it is
      even the identity for $b = b_\r$;
    \item[(IN-2)] for every $a \in \dom(T)$, $\restr{\pi'}{\ineigtwo{T}{a}}$ is an
      isomorphism between $\ineigtwo{T}{a}$ and $\ineigtwo{\calJ}{\pi'(a)}$;
    \item[(Surj-P)] $\phi$ is \deft{surjective} for proper bags except for
      $\calW$: for any
      proper bag $b$ of $\calJ$ whose domain is not a subset of $\calW$, $b$ has a
      preimage by $\phi$;
    \item[(FD-S)] $T$ (seen as an interpretation) is FD-safe except for $\calW$;
    \item[(Achieve)] for any $a \in \dom(T)$, for any relation $R$ of $\sigma_{>2}$, for any
      subset $P$ of the positions of $R$ which is not a strict superset of an FD
      determiner of $\Phi$, if $\pi(a)$ is such that $(t', \pi(a)) \in
      R_i^{\calJ}$ for some $t'$ for all $R^i \in P$, then the same is true of $a$ in $T$
      (seen as an interpretation) for some
      $t$. Further, unless $P$ is exactly an FD determiner, letting $S \defeq
      \ineig{T}{t} \backslash \{R_i(t, a) \mid R^i \in P\}$ and $S'  \defeq
      \ineig{\calJ}{t'} \backslash \{R_i(t', \pi(a)) \mid R^i \in P\}$,
      $\restr{\pi}{S}$ is an isomorphism between $S$ and $S'$.
  \end{description}
\end{definition}

Intuitively, property (Achieve) is designed to preserve exactly what can be
asserted by non-conflicting rules. Except in the case where the
frontier variables are exactly a determiner, this includes the patterns of
equalities between the ``non-frontier'' variables of the atom. We cannot preserve
more, because we need to remain FD-safe.

We modify the definition of unravelable interpretations to require that all elements
of the interpretation occur in at least one tuple for a binary relation not in $\sigmas \backslash
\sigma$, and that its bag graph is connected even when the non-proper bags are
removed. This can be ensured without loss of generality as before, because
the fresh binary relation used to ensure the condition is not in $\sigmas
\backslash \sigma$.

We must show the correctness of such unravelings:

\begin{proposition}
  \label{prp:unravelfdok}
  For any $\sigmas$-fact $F$, $\gctwo$ constraints $\Sigma$, CQ $q'$, FDs $\Phi$,
  and non-conflicting $\fone$ constraints $\Delta$,
  if $\calJ$ is an unravelable interpretation 
  that satisfies $\Sigma$, $\shred(\Delta)$, $\wf(\sigmas)$, $\neg q'$, and has
  a witness $\calW$ of $F$, and
  $T$ is an FD-faithful unraveling of $\calJ$ preserving $\calW$, then $T$ is an
  interpretation which is FD-safe except for $\calW$, it satisfies 
  $\Sigma$, $\shred(\Delta)$, $\wf(\sigmas)$, and $\neg q'$, and $\calW$ is
  still a witness of $F$ in $T$.
\end{proposition}

\begin{proofb}
  $T$ is clearly FD-safe except for $\calW$ by property (FD-S), and it satisfies $\neg q'$ (by the
  homomorphism $\pi$). It satisfies $\Sigma$ and $\wf(\sigmas)$ by the same arguments as in the
  proof of Proposition~\ref{prp:constraintsok}, noting that $\Sigma$ and
  $\wf(\sigmas)$ do not
  refer to the fresh relations of $\sigmas$, so it is sufficient to have
  isomorphisms between arity-two immediate neighborhoods, and to have
  surjectivity of $\pi$ for the proper bags only. The harder part is to show that
  $\shred(\Delta)$ is satisfied.

  Consider any $\tau \in \Delta$, and consider a match $f$ of the body of
  $\shred(\tau)$ in $T$, and let $a$ be the element of $\dom(T)$ to which the
  frontier variable of
  $\tau$ is mapped.
  Consider the image of $f$ by the homomorphism $\pi$ in $\calJ$.
  As $\calJ$ satisfies
  $\shred(\tau)$, this implies that the element $a' \defeq \pi(a)$ in
  $\dom(\calJ)$ is
  such that the head of $\shred(\tau)$ can be matched to~$\calJ$ with a homomorphism
  mapping the frontier variable to~$a'$. Now $\tau$ is a single-head dependency,
  and we made the assumption that heads were
  either unary or higher-arity. If the head of $\tau$ is unary, then, so is the
  head of $\shred(\tau)$, and,
  considering the 
  restriction of $\pi$ to any proper bag containing $a$ in $T$
  (such a bag exists as we assumed that the interpretation is unravelable), as this
  restriction is an isomorphism, we conclude that the unary head atom to which
  the head of $\shred(\tau)$ is matched in $\calJ$ also has a match in $T$, so that $f$ does not
  witness a violation of $\shred(\tau)$. Hence, let us assume that the head of $\tau$ is
  higher-arity, and let $R$ be the higher-arity relation.

  This means that there is a subset $P$ of positions of $R$ (namely, the set of
  positions of the head of $\tau$ where the frontier variable occurs), and there
  is $t' \in \dom(\calJ)$, such that $(t', a') \in R_i^{\calJ}$ for all $R^i \in P$.
  We know by the non-conflicting condition that $P$ is not a strict
  superset of a determiner of an FD in $\Phi$. If $P$ is exactly a determiner of
  an FD in $\Phi$, property (Achieve) ensures $(t, a) \in R_i^T$ for all $R^i \in P$
  for some $t \in \dom(T)$. Now, by the non-conflicting condition, all variables
  in the head of $\tau$ at positions not in $P$ are existential variables and it
  is their only occurrence. Hence, the fact that $T$ satisfies $\wf(\sigma)$
  ensures that the head of $\shred(\tau)$ has a match in $T$ mapping the
  frontier variable to~$a$, so that $f$ does not witness a violation of $\shred(\tau)$.

  If $P$ is not  a determiner of an FD, then property (Achieve) ensures that
  $(t, a) \in R_i^T$ for all $R^i \in P$
  for some $t \in \dom(T)$ and $\ineig{T}{t}
  \backslash \{R_i(t, a) \mid R^i \in P\}$ and $\ineig{\calJ}{t'} \backslash
  \{R_i(t', a') \mid R^i \in P\}$ are
  isomorphic. This implies that the head of $\shred(\tau)$ has a suitable match in $T$
  so that $f$ does not witness a violation of $\shred(\tau)$; indeed, seeing the
  tuples $(t, a'')$ and $(t', a''')$ in $R_i^T$ and $R_i^{\calJ}$ as ground
  $R$-atoms $A_1$ and $A_2$,
  the head atom $A$ of
  $\tau$ has a homomorphism to $A_2$ mapping the frontier variable to $a'$, and we know that the elements at positions
  of $A_1$ and $A_2$ which are not in $P$ have the same equalities, and that $A$
  contains the frontier variable at positions $P$ and other
  variables at the other positions; so $A$ also has a homomorphism to $A_2$
  mapping the frontier variable to $a$. Hence, $T$
  satisfies $\Delta$.

  This justifies that $T$ satisfies all the required constraints, concluding the
  proof.
\end{proofb}

We now describe the FD-faithful unraveling process:

\begin{proposition}
  \label{prp:haveunravelfd}
  For any fact $F$,
  for any set $\Phi$ of FDs on $\sigma_{>2}$,
  for any unravelable interpretation $\calJ$ of $\wf(\sigmas)$ and witness $\calW$ of $F$ in $\calJ$
  such that the unshredding of $\calW$ satisfies $\Phi$,
  there is an FD-faithful unraveling $T$ of $\calJ$ preserving $\calW$.
\end{proposition}

\begin{proofb}
  We modify the proof of Proposition~\ref{prp:haveunravel} in the two ways.

  The first modification is that, whenever we unravel on a bag $b$ where the element $a$ was introduced, and
  $(\pi(a)) \in \elt^{\calJ}$,
  we deal differently with the non-proper bags adjacent to
  $\phi(b)$ in the bag graph of $\calJ$. We now give details.

  Let $\calB$ be the set of non-proper bags in the bag graph of $\calJ$ that share
  $a$ with $\phi(b)$, to which we add $\phi(b)$ itself if it is non-proper. We consider all subsets $P$ of positions of all
  higher-arity relations $R$ such that $(t', \pi(a)) \in R_i^{\calJ}$ for some
  $t'$ for all $R^i \in P$, and $P$ is not a strict superset of a determiner of
  an FD: we say that $\pi(a)$ \deft{occurs} at~$P$. For any such $P$, we say
  that $b'$ (necessarily in $\calB$) \deft{realises} $P$ if $b'$ witnesses that
  $\pi(a)$ occurs at~$P$. We add the following children (for non-proper bags;
  for proper bags we do as before): for every such $P$
  which is not an FD determiner, for every bag $b'$ of $\calB$ that realizes
  $P$, create one child of $b$ for~$b'$ containing the tuples that witness that $\pi(a)$
  occurs at~$P$, and unravel on this child; for every such $P$ which is an
  FD determiner, and for which it not already the case that $(t, a) \in R_i^T$ for
  some $t$ for all $R^i \in P$, pick \emph{one} bag $b'$ of $\calB$ that
  realizes $P$, create one child of $b$ for~$b'$ containing the tuples that
  witness that $\pi(a)$ occurs at~$P$, and unravel on this child.

  In other words, informally, for the non-proper bags, we look at all sets of
  positions of higher-arity relations in which $\pi(a)$ occurs, keeping only those
  which are not a strict superset of an FD determiner. For those which are not
  FD determiners, we trigger-happily unravel on every bag where $\pi(a)$ occurs
  at these positions. For those which are FD determiners, we only unravel if $a$ does no
  already occur at those positions, and then we choose only one representative
  bag. In all cases, if the current bag $\phi(b)$ is non-proper, we also include
  it in the bags that we examine: this may mean that we have an infinite chain
  in $T$ of copies of this bag, but this is not a problem, as $T$ is infinite.
  Also, in all cases, when unraveling on a non-proper bag, we only copy the
  tuples witnessing that $\pi(a)$ occurs at the relevant positions; if $\pi(a)$
  occurred at other positions, we do not copy such tuples (we will complete the
  other positions when unraveling at the next step, see below).

  The second modification is that, when we unravel on a bag $b$ where the element $t$ was introduced (and
  the other element is $a'$), and we have
  $(\pi(t)) \notin \elt^{\calJ}$ (so that, as $\calJ$ satisfies
  $\wf(\sigmas)$, $(\pi(t)) \in A_R^{\calJ}$ for some $R \in \sigma_{>2}$), we
  compare the tuples of $b$ and of $\phi(b)$. Indeed, by the first modification,
  it may be the case that some tuples of $\phi(b)$ were not copied in $b$.
  We let $P'$ be the positions of
  $R$ such that $(t, a') \notin R_i^b$ but $(\pi(t), \pi(a')) \in R_i^{\calJ}$.
  In addition to the neighbors of $\phi(b)$ in the bag graph that
  we would ordinarily consider, we consider a ``virtual'' neighbor, a bag
  containing the tuples $(\pi(t), \pi(a'))$ in its interpretation of $R_i$ for~$R^i \in P'$, on which we
  also unravel as usual.

  In other words, informally, when unraveling non-proper bags representing a
  higher-arity ground atom where one
  element of the atom, introduced at the parent bag of~$b$ in $T$, occurs at multiple
  positions, and we have only kept a subset of these positions, then the missing
  occurrences are seen as another bag, that we will also copy (and which may make us go
  back, in $G$, to the bag for the parent of $b$ in $T$).

  To give an example, assume that $(u) \in U^{\calJ}$, we are unraveling on
  some element $u$, and $\calJ$
  contains the shredding of the $R$-tuples $(u, u, u, u, v)$ and $(u, u, v, v,
  v)$. Say $R^1 R^2$ and $R^2 R^3$ are the determiners of the FDs in $\Phi$ on
  $R$. The unraveling will
  create the shredding of the following ground atoms:
  \begin{itemize}
    \item for $R^1$, the $R$-tuples $(u, u_1, u_1, u_1, v_1)$ and $(u, u_2, v_2,
      v_2, v_2)$ -- note how the other positions where $u$ occurs contain a
      fresh copy of $u$, created when unraveling on the virtual neighbor;
    \item for $R^2$, the $R$-tuples $(u_3, u, u_3, u_3, v_3)$
      and $R(u_4, u, v_4, v_4, v_4)$;
    \item for $R^3$, the $R$-tuple $(u_5, u_5, u, u_5, v_5)$;
    \item for $R^4$, the $R$-tuple $(u_6, u_6, u_6, u, v_6)$;
    \item for $R^1 R^2$, the $R$-tuple $(u, u, u_7, u_7, v_7)$ -- note that only the first
      tuple was used as witness, and indeed using also the second would have violated
      FD-safety;
    \item for $R^2 R^3$, the $R$-tuple $R(u_8, u, u, u_8, v_8)$.
  \end{itemize}

  We show correctness. Properties (Compat-P), (IN-2) hold for the same reasons
  as in the original construction, and (Surj-P) holds
  because the Gaifman graph of $\calJ$ is connected using a fresh
  binary relation which we consider as a proper bag. FD-safety (property
  (FD-S)) holds initially
  because the copy of the witness $\calW$ contains no shredding of higher-arity
  tuples except the ones that occur in the unshredding of
  $\calW$, which satisfies $\Phi$.
  We show that the property is preserved during the unraveling, by observing that,
  whenever we create atoms of $\sigmas \backslash \sigma$, of the form $(t,
  a)$ for a relation $R_i$, for $R^i$ in a set $P$ of positions of $R$, then either $a$ is
  fresh in $T$, or $P$ does not contain an FD determiner, or $P$ is an FD
  determiner but $a$ does not occur at these positions already.

  We now check that property (Achieve) is satisfied. For any $a \in \dom(T)$,
  consider the bag $b$ of $T$ where $a$ was introduced, and let us check the
  condition. The first part of the condition is clear by our construction: for
  any such $P$, there is a child of $b$ witnessing that $a$ occurs at the right
  positions in $T$. The second part of the condition holds because, when $P$ is
  not an FD determiner, we create one child for each bag that realizes $P$, and
  unravel on this child.
\end{proofb}

We conclude as in Lemma~\ref{lem:unravel}: we make the original interpretation unravelable
without loss of generality, we apply Proposition~\ref{prp:haveunravelfd}, and
conclude by Proposition~\ref{prp:unravelfdok} and Lemma~\ref{lem:cfree}.

\end{document}